\documentclass[fleqn,usenatbib]{mnras}

\usepackage{newtxtext,newtxmath}
\usepackage[T1]{fontenc}

\DeclareRobustCommand{\VAN}[3]{#2}
\let\VANthebibliography\thebibliography
\def\thebibliography{\DeclareRobustCommand{\VAN}[3]{##3}\VANthebibliography}

\usepackage{graphicx}	% Including figure files
\usepackage{xcolor}

\title[DM interactions and JWST ]{The feasibility of constraining DM interactions with high-redshift observations by JWST}

\author[A. Kurmus et al.]{Ali Kurmus$^{1}$,\thanks{E-mail: alikurmus@college.harvard.edu}
Sownak Bose$^{2,1}$,\thanks{E-mail: sownak.bose@durham.ac.uk}
Mark Lovell$^{3}$, 
Francis-Yan Cyr-Racine$^{4}$,
Mark Vogelsberger$^{5}$,
\newauthor
Christoph Pfrommer$^{6}$,
and 
Jes{\'u}s Zavala$^{3}$
\\
$^{1}$Center for Astrophysics, Harvard \& Smithsonian, 60 Garden St, Cambridge, MA 02138, USA\\
$^{2}$Institute for Computational Cosmology, Durham University, South Road, Durham DH1 3LE, UK\\
$^{3}$Center for Astrophysics and Cosmology, Science Institute, University of Iceland, Dunhagi 5, 107 Reykjavik, Iceland\\
$^{4}$Department of Physics and Astronomy, University of New Mexico, 210 Yale Blvd NE, Albuquerque, NM 87106\\
$^{5}$Department of Physics, Kavli Institute for Astrophysics and Space Research, Massachusetts Institute of Technology, Cambridge, MA 02139, USA\\
$^{6}$Leibniz-Institute for Astrophysics Potsdam (AIP), An d. Sternwarte 16, 14482 Potsdam, Germany
}

\date{Accepted XXX. Received YYY; in original form ZZZ}

\pubyear{2022}

\begin{document}
\label{firstpage}
\pagerange{\pageref{firstpage}--\pageref{lastpage}}
\maketitle

\begin{abstract}
Observations of the high redshift universe provide a promising avenue for constraining the nature of the dark matter (DM). This will be even more true with the advent of the {\it James Webb Space Telescope} (JWST). We run cosmological simulations of galaxy formation as part of the {\it Effective Theory of Structure Formation} (ETHOS) project to compare high redshift galaxies in Cold (CDM) and alternative DM models which have varying relativistic coupling and self-interaction strengths. The interacting DM scenarios produce a cutoff in the linear power spectrum on small-scales, followed by a series of `dark acoustic oscillations'. We find that DM interactions suppress the abundance of galaxies below $M_\star \sim 10^8\,{\rm M}_\odot$ for the models considered. The cutoff in the power spectrum delays structure formation relative to CDM. Objects in ETHOS that end up at the same final masses as their CDM counterparts are characterised by a more vigorous phase of early star formation. While galaxies with $M_\star \lesssim 10^6\,{\rm M_\odot}$ make up more than 60 per cent of star formation in CDM at $z\approx 10$, they contribute only about half the star formation density in ETHOS. These differences diminish with decreasing redshift. We find that the effects of DM self-interactions are negligible compared to effects of relativistic coupling (i.e. the effective initial conditions for galaxy formation) in all properties of the galaxy population we examine. Finally, we show that the clustering strength of galaxies at high redshifts depends sensitively on DM physics, although these differences are manifest on scales that may be too small to be measurable by JWST.
\end{abstract}


%%%%%%%%%%%%%%%%%%%%%%%%%%%%%%%%%%%%%%%%%%%%%%%%%%

%%%%%%%%%%%%%%%%% BODY OF PAPER %%%%%%%%%%%%%%%%%%

\section{Introduction}

Dark Matter (DM) plays a significant role in structure formation in the Universe, but the exact nature of this building block is still uncertain. The current paradigm assumes that DM is cold and collisionless,  corresponding to the so-called Cold DM (CDM) \citep{Blumenthal_1984}. The CDM model is very successful at describing the distribution of matter on large scales and at predicting properties of the galaxy population across cosmic time \citep{Davis_1985, Frenk_1985, White_1987, White_1987b, Frenk_1988, Frenk_1990, Bertschinger_1991, Springel_2005}. A review of the progress of DM theories can be found in \citet{Frenk_2012}.

Despite its successes, comparisons of the CDM with existing observations have shown discrepancies at the dwarf galaxy scale. Among the small-scale ``problems'', the most prominent ones are the ``Missing Satellites'' (MS), which refers to the excess abundance of dwarf galaxies predicted by CDM model compared to observations \citep{Klypin_1999, Moore_1999},  the ``Too Big to Fail'' \citep{BoylanKolchin_2011} which concerns the fact that observationally inferred circular velocities of subhaloes hosting the largest satellites of the Milky Way are lower than the velocities of the most massive subhaloes in simulations, these subhaloes should be too big to fail to form galaxies, the ``Cusp-core'' \citep{Flores_1994, deBlok_2001} which states that the centres of the haloes predicted by the CDM model are too ``cuspy'' and dense compared to observations, and the ``Planes of Satellites'' \citep{Ibata_2014, Pawlowski_2014}, which argues that the observed thin planar configurations of galactic satellites appear quite rarely in simulations. \citet{Bullock_2017} presents a comprehensive review of these dicrepancies. However, these issues may be resolved through a more comprehensive understanding of baryonic physics and galaxy formation processes. While many of these problems were identified in DM-only simulations, a more sophisticated treatment of baryonic physics alleviates the CDM problems without the need of considering different particle physics alternatives \citep{Kauffmann_1993, Pontzen_2012, Brooks_2013, DiCintio_2014, Cautun_2015, Sawala_2016, Read_2017, Kim_2018, Genina_2018}. For example, modern hydrodynamical simulations show that stellar and radiative feedback suppress star formation, which results in DM haloes that do not host a galaxy. This effect provides a natural solution to the ``missing satellites'' problem. Furthermore, supernova feedback in dwarf galaxies can result in core formation by heating the inner DM distribution of the halo, which will convert a ``cuspy'' halo to a one with a central core. 

 While there remain some open questions regarding the full efficacy of baryonic solutions to each of the small-scale challenges listed above, what remains uncontroversial is the lack of evidence for a CDM particle. The main motivator, therefore, for alternative models is the absence of a detection of the weakly interacting massive particles (WIMPs) proposed by the simplest models of the CDM theory, rather than necessarily any problems associated with the predictions it makes. Despite thorough searches at direct-detection and accelerator-based facilities, WIMPs remain undetected \citep{Lanfranchi_2020}. In addition, the absence of clear signatures of supersymmetry at the Large Hadron Collider provides further motivation to consider DM candidates beyond the WIMP \citep{Barman_2020}. 

So far, searches for the particle that forms the basis of CDM did not yield any positive results. This lead to the development of other well-motivated DM models that break two of the primary assumptions of the CDM model: DM being cold and DM being collisionless. Modifications to these basic properties lead to a difference in the non-linear structure formation process which could be a solution to small scale challenges of the CDM model. The alternative models are generally based on modifying the small-scale fluctuations in the CDM model while keeping its success at describing large scales. The two most prominent alternative models that separately relaxed the basic assumptions of CDM include ``warm dark Matter'' (WDM) and ``self-interacting dark matter'' (SIDM).

In the WDM model DM particles were near relativistic in the early universe which resulted in their free streaming and smoothed out density perturbations below some characteristic scale. This property is reflected as a cutoff in the initial matter power spectrum and suppresses the formation of low mass structures \citep{Maccio_2010, Kennedy_2014, Bose_2017, Bozek_2019}. A leading candidate for WDM is the resonantly-produced sterile neutrino \citep{Asaka_2005, Abazajian_2006, Boyarsky_2009, Adhikari_2017}. Compared to WIMPs, whose annihilation products could be detected as $\gamma$ radiation, the energy range for detecting these particles is in the X-ray regime. Several papers have reported a detection of a candidate X-ray decay line (using XMM -Newton data) in galaxies and galaxy clusters, for which the DM particle has a mass of 7.1~keV \citep{Bulbul_2014, Boyarsky_2014}. However these detections were not associated with the DM candidate after non-detections that followed \citep{Jeltema_2014, Anderson_2015, Dessert_2021}. Neither the Chandra nor Hitomi X-ray satellites \citep{Hofmann_2016, Aharonian_2017} nor the blank observations \citep{Dessert_2020} observed such a spectral feature. This strongly challenged the DM interpretation and raised questions about uncertainties in the X-ray detector responses. %This means neutrinos could be detected using less energy.

On the other hand, SIDM is a model where, unlike in CDM, DM particles are no longer collisionless, and may scatter upon interaction. One of the original motivations of the model was that with large enough self-interaction cross-sections per unit mass, SIDM could modify both the abundance and internal structure of haloes and subhaloes \citep{Spergel_2000}. Later, \citet{Miralda_2002} argued that the proposed cross-sections were too large and SIDM could not be a plausible DM candidate. \citet{Peters_2013} relaxed the constraints on the cross-section limit by an order of magnitude. However, DM interactions alone as a modification was still not sufficient to address all the CDM discrepancies \citep{Zavala_2013,Brooks_2014}. Although a DM model with only self-collisions was not strong enough to have an impact on the power spectrum on presently measurable scales, modifications to SIDM such as interactions with a Yukawa-like potential \citep{Vogelsberger_2012} or inelastic self-interactions \citep{Vogelsberger_2019} are able to largely address any noted discrepancies in the CDM model (i.e. without any extra baryonic physics). Specifically, a DM model that couples with relativistic particles in the early universe, in addition to self collisions, is able to suppress the formation of dwarf scale haloes \citep{Boehm_2002, Buckley_2014, Boehm_2014}. Thus, by allowing the vector or scalar field mediating the DM self collisions to couple with sterile neutrinos or any other form of dark radiation (DR), this idea of relativistic interactions can also address the ``Too Big To Fail'' and ``cusp-core'' problems of the CDM \citep{VandenAarssen_2012, Bringmann_2014, Dasgupta_2014}. On galaxy scales, the particle physics modification is a cutoff in the initial power spectrum similar to WDM. However, unlike the smooth monotonic power law cutoff in the case of WDM, this modification results in more complex behaviour after the cutoff such as dark acoustic oscillations and Silk damping tail \citep{Buckley_2014}.

For many DM models, the shape and the amplitude of the initial matter power spectrum, and the velocity dependence of the self scattering of DM particles significantly determine both the abundance and structure of the DM haloes on a variety of mass scales. Hence, DM models that make similar predictions to power spectrum and cross section can be classified under a single ``effective theory of structure formation'' (ETHOS) \citep{Cyr-Racine_2016}. This is the framework within which we investigate our present work. We provide more details about the origin and derivation of ETHOS in Section \ref{sec:Dark_Matter}. 

Previous cosmological simulations that have compared CDM with models that had a cutoff in the power spectrum showed that the most prominent differences between the models were in the early universe \citep{Dayal_2015, Lovell_2018, Lovell_2019, Khimey_2020}. With the launch of the {\it James Webb Space Telescope} (JWST), the high redshift universe will be accessible with significantly better statistics than previously. Hence, the launch of the JWST provides an incredible opportunity to push constraints on DM into a thus far uncharted regime, enabling us to derive stronger constraints than ever before. 

The goal of this work is to quantify, using state-of-the-art hydrodynamical simulations, whether the differences in the power spectra would lead to observable differences by JWST, and if they do, to evaluate whether these could be mapped to the difference in initial conditions. We compare two extreme ETHOS models with CDM, in which the particle physics parameters have been chosen so as to yield prominent differences in the observable scale of galaxies. One of these models has self-interactions enabled. Our objective is to use these observable differences to disentangle how these DM interactions manifest onto galaxy scales. We also compare a model that was designed specifically for the purpose of addressing the small scale challenges to CDM described above to ascertain whether this model generates observed deviations from CDM at high redshift.

This paper is organised as follows. We present the different ETHOS models we compare with CDM in detail, by showing the exact shapes of their power spectra, strength of coupling, and extent of self collisions, in Section~\ref{sec:Dark_Matter}. In Section~\ref{sec:Simulations}, we describe the hydrodynamical simulations we use and provide our initial conditions. We then present our results for the overall statistics of the galactic populations in these models in Section~\ref{sec:Results}, and present a first calculation of the strength of galaxy clustering in models of these kinds in Section~\ref{sec:clustering}. Finally, we summarise our conclusion in Section~\ref{sec:Conclusions}.

\section{DARK MATTER MODELS}
\label{sec:Dark_Matter}

In this work, the non-CDM models we consider are those in which a massive fermionic DM particle couples with a relativistic species (neutrinos, dark radiation) via a massive mediator. The origin and derivation of such models are described in detail in \citet{Cyr-Racine_2016} within the ETHOS framework and their first direct comparison with CDM in terms of structure evolution was made in \citet{Vogelsberger_2016}. Furthermore, \citet{Bohr_2020} presents a simple parametrization of the power spectra of these type of models using the scale and the amplitude of the first DAO peak and \citet{Munoz_2020} uses such parametrization to discuss the effects of ETHOS models on 21-cm hydrogen line. The ETHOS framework receives particle physics parameters as an input, and enables us to map these quantities into the linear power spectra and (velocity-dependent) self-interaction cross-sections corresponding to the DM particle. 

Each model is primarily characterised by the strength of the DM coupling with the relativistic species and by the amount of interactions among DM particles (which we hereafter refer to as DM-DR, which denotes dark radiation, and DM-DM interactions respectively). The standard CDM model does not have relativistic interactions or self-interactions. One of the models we consider, dubbed ETHOS4, is a model within the ETHOS framework where the DM-DR and DM-DM interactions have been fine-tuned with the sole purpose of alleviating the Missing Satellites and Too Big To Fail problems through DM physics alone. Its characterising parameters are the combination of DM-DR and DM-DM interactions which would best solve CDM problems. The details of the model can be found in \citet{Vogelsberger_2016}.

The other two models of comparison in this work are ``atomic'' DM models with strong relativistic interactions. They are introduced and further explained in \citet{Kaplan_2010, Kaplan_2011, Cyr-Raccine_2013, Bose_2019}, and are referred to as sDAO and noSIDM. The former includes the effects of DM self-interactions, while the latter does not. sDAO and noSIDM are again models involving the coupling of DM particles with relativistic species. Unlike ETHOS4, they were not created for only solving the CDM problems. Rather, they serve as a useful benchmark for quantifying the observable impacts of a DM model where the primordial differences relative to CDM are especially prominent.
\begin{figure}
    \centering
    \includegraphics[width=\columnwidth]{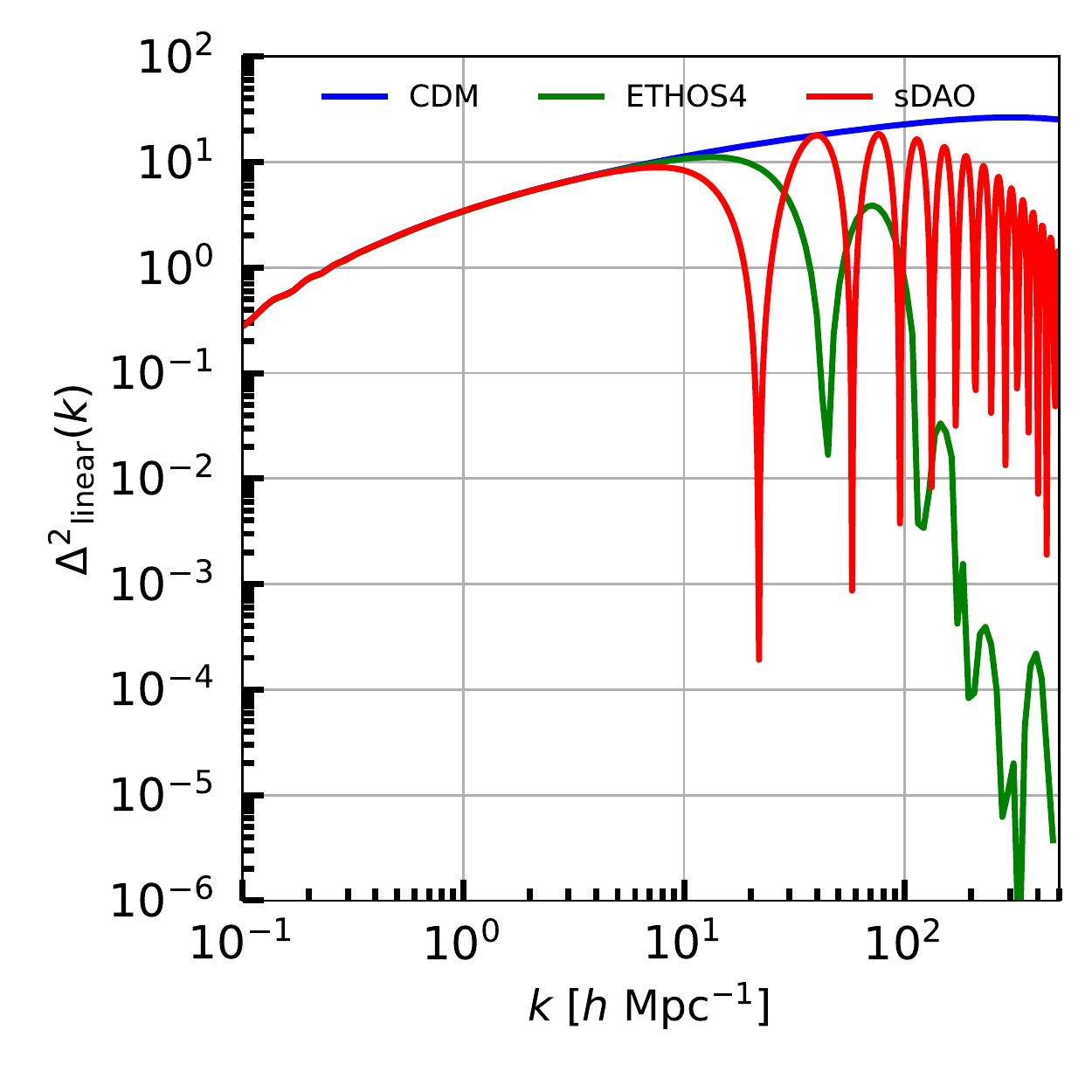}
    \caption{Dimensionless linear matter power spectra $\Delta ^2_{\rm linear}(k) = k^3 P_{\rm linear}(k)/2\pi^2$ as a function of comoving wave number $k$ for CDM, ETHOS4, and sDAO/noSIDM. Larger wavenumbers corresponds to smaller-scale structure. The alternative models differ from CDM at around dwarf scales. Compared to CDM, ETHOS models have a cutoff at the power spectra and oscillations after the cutoff at smaller scales. The number of undamped acoustic oscillations depends on the strength of coupling with relativistic particles. sDAO has the most frequent and highest amplitude oscillations and the greatest cutoff scale. As the colour shifts from blue to red, the models deviate more extremely from CDM on particle physics level.}
    \label{fig:power_spectra}
\end{figure}

Figure~\ref{fig:power_spectra} compares the linear power spectra in each of these models. We see that the sDAO model has a cutoff in the power spectrum at larger scales and has more pronounced ``dark acoustic oscillations'' (DAOs) compared to ETHOS4. The shape of the power spectrum depends on how strongly DM couples with relativistic particles and how quickly it decouples from them. A strong coupling that ends at very high redshift would not cause a significant difference in structure formation compared to CDM just as a very weak coupling would not. From a more quantitative perspective, around the redshift of DM kinetic decoupling, $z_{\rm D}$, we approximately have:
\begin{equation}
\label{eq:w_vs_s_DAO}
(\dot{\kappa}_\chi/\mathcal{H})|_{z \sim z_{\rm D}}\simeq (z/z_{\rm D})^n,    
\end{equation}
where $\dot{\kappa}_\chi$ is the interaction rate between DM and relativistic species and $\mathcal{H}$ is the conformal Hubble expansion rate \citep{Bose_2019}. The $\dot{\kappa}$ indicates a derivative with respect to conformal time. Generally, a larger value of $n$ results in greater amounts of DAOs on the smaller scales of the power spectra. Values of $n\lesssim4$ result in structure formation comparable to standard WDM models, while values of $n\gtrsim6$ result in a significantly different structure formation (which defines the general class of strong dark acoustic oscillation, or sDAO, models, see also \citet{Bohr_2020}).

In sDAO models, the coupling between the DM and relativistic species are substantially stronger than in ETHOS4. Within the ETHOS framework, the parameters of the ETHOS4 and sDAO models are respectively:
\begin{equation*}
    \left\{ n,a_n,\omega_{{\rm DR}}, \alpha_{l\geq2} \right\} = \left\{ 4\,, 414\, {\rm Mpc}^{-1}, 1.35\times10^{-6},3/2 \right\}
\end{equation*}
 and
\begin{equation*}
    \left\{ n,a_n,\omega_{{\rm DR}}, \alpha_2, \alpha_{l\geq3} \right\} = \left\{ 6\,, 6\times10^{8} {\rm Mpc}^{-1}, 1.25\times10^{-8},9/10\,,1 \right\}
\end{equation*}
where $n$ is the exponent in Equation~\ref{eq:w_vs_s_DAO}, $a_n$ is the normalization of the interaction rate between DM and relativistic particles, $\omega_{\rm DR}\equiv\Omega_{{\rm DR}}h^2$ is the physical energy density of the dark radiation in terms of critical density, and $\alpha_l$ is a set of coefficients that defines the angular dependence of the DM-dark radiation scattering cross section \citep{Cyr-Racine_2016,Bose_2019}.

The sDAO models considered in this work are already constrained by current observations \citep{Viel_2013, Irsic_2017}. Our aim is to maximise these effects in any case, in an effort to find out whether the DAOs will result in differences that would be observable by JWST and, if they do, whether the cause of the difference could be clearly mapped to DM-DR  or DM-DM interactions. Table \ref{tab:dm_models} summarizes the properties and purposes of the different models used in this work.

\begin{table}
	\centering
	\caption{Summary of the properties and purpose of different DM models.``Optimal'' in the table means fine tuned to exactly alleviate the MS and TBTF problems in CDM.}
	\label{tab:dm_models}
	\begin{tabular}{lccr} % four columns, alignment for each
		\hline
		Model & DM-DR & DM-DM & Purpose\\
		\hline
		CDM & - & - & Standard Base Model\\
		ETHOS4 & optimal & optimal & Alleviates MS and TBTF \\
		noSIDM & strong & - & Tests sDAOs \\
		sDAO & strong & strong & Tests sDAOs and Self-Interactions\\
		\hline
	\end{tabular}
\end{table}

\section{Simulations}
\label{sec:Simulations}

The simulations used in this work comprise the CDM and ETHOS4 simulations introduced in \citet{Lovell_2018} plus two new simulations -- of sDAO and noSIDM -- that were run for this study. For all simulations, we use a modified version of the {\sc Arepo} code \citep{Springel_2010}, augmented by with the IllustrisTNG galaxy formation model \citep{Torrey_2014, Vogelsberger_2014a, Vogelsberger_2014b, Genel_2014, Weinberger_2017, Pillepich_2018}. Similarly to \citet{Lovell_2018} we use a significantly extended version of {\sc Arepo} which includes isotropic and elastic DM-DM interactions \citep{Vogelsberger_2016} following arbitrary velocity-dependent interaction cross sections. All four simulations share the cosmological parameters of: DM density $\Omega_{0}=0.302$, dark energy density
$\Omega_{\Lambda}=0.698$, baryon density $\Omega_{\rm b}=0.046$, Hubble
parameter $H_0 = 100\,h\,{\rm kms}^{-1} {\rm Mpc}^{-1} = 69.1\,{\rm kms}^{-1} {\rm Mpc}^{-1}$,  power
spectrum normalisation $\sigma_{8}=0.838$, and power spectrum slope index
$n_\rmn{s}=0.967$. All initial conditions have been created using the same initial phases. We perform the simulations over a box volume of $(36.2\, {\rm Mpc})^3$. The DM particle mass, comoving DM softening length, average gas cell mass, and comoving minimum gas softening length are $1.76\times10^{6} {\rm M}_\odot$,  $724$ pc, $2.69 \times 10^5 {\rm M}_\odot$, and $181$ pc, respectively. The simulations are evolved from $z = 127$ to $z = 6.5$. We identify haloes in the simulations using the Friends-of-Friends (FoF) algorithm, and determine the gravitationally bound subhaloes using the {\sc Subfind} algorithm \citep{Springel_2001}.

\section{Results}
\label{sec:Results}
In this work, we examine the differences in simulated universes of different DM models varying in strength of DM-DR and DM-DM interactions. These two quantities affect the abundance and internal structure of DM haloes. However, a mere difference in haloes will not help to observationally constrain DM models. Thus, we further focus on the implications of the halo differences on the observable aspects, specifically those potentially measurable in the era of JWST. To this end, we focus on the differences in the predicted galaxy properties between models at high redshift. In order to gain intuition for how the different particle physics processes affect the growth of structure, we begin by comparing the statistics of their respective DM halo populations.

\subsection{Halo mass functions (HMF)}
\label{sec:HMF}
\begin{figure}
\centering
    \includegraphics[width =\columnwidth]{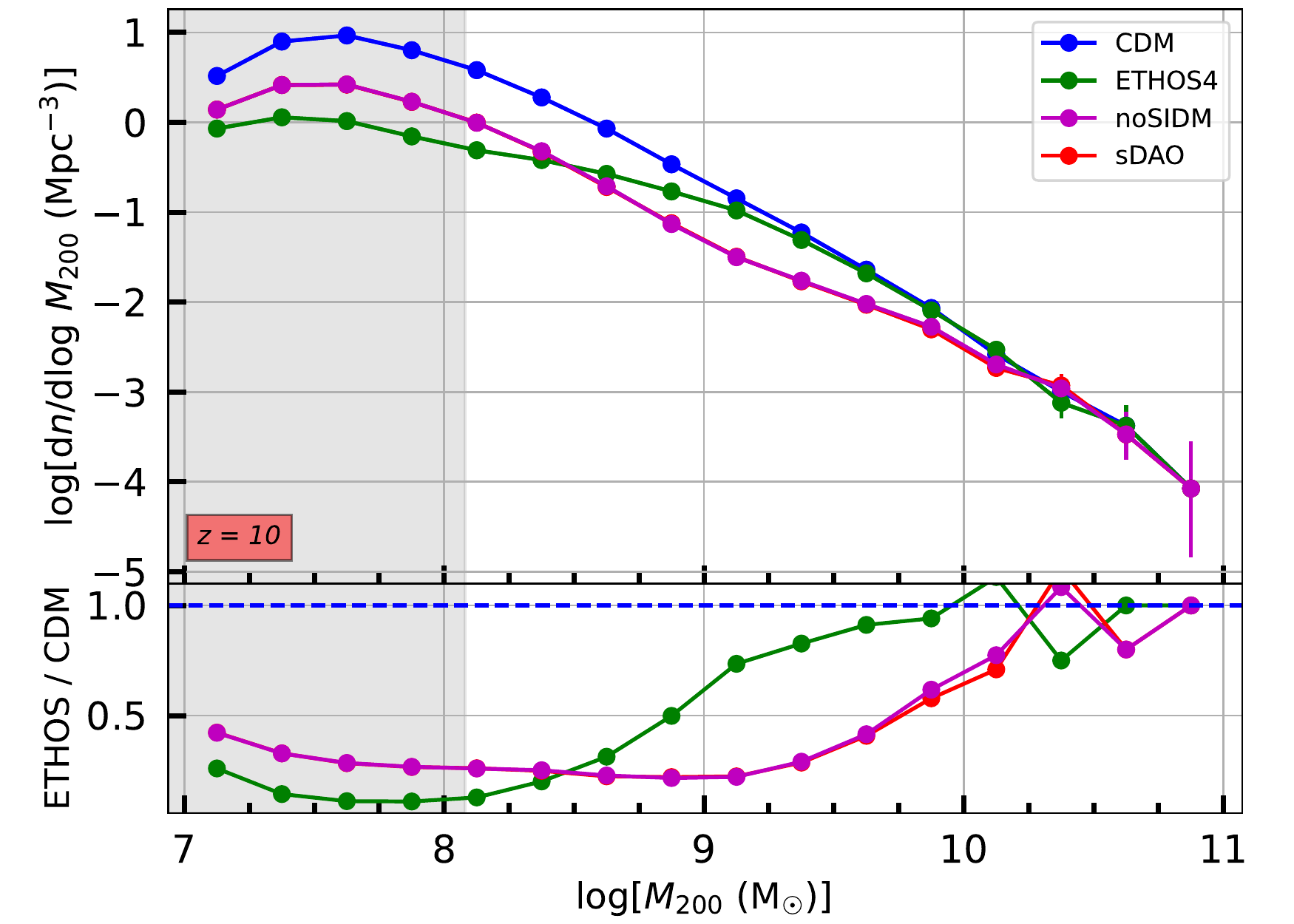}
    \\
    \includegraphics[width =\columnwidth]{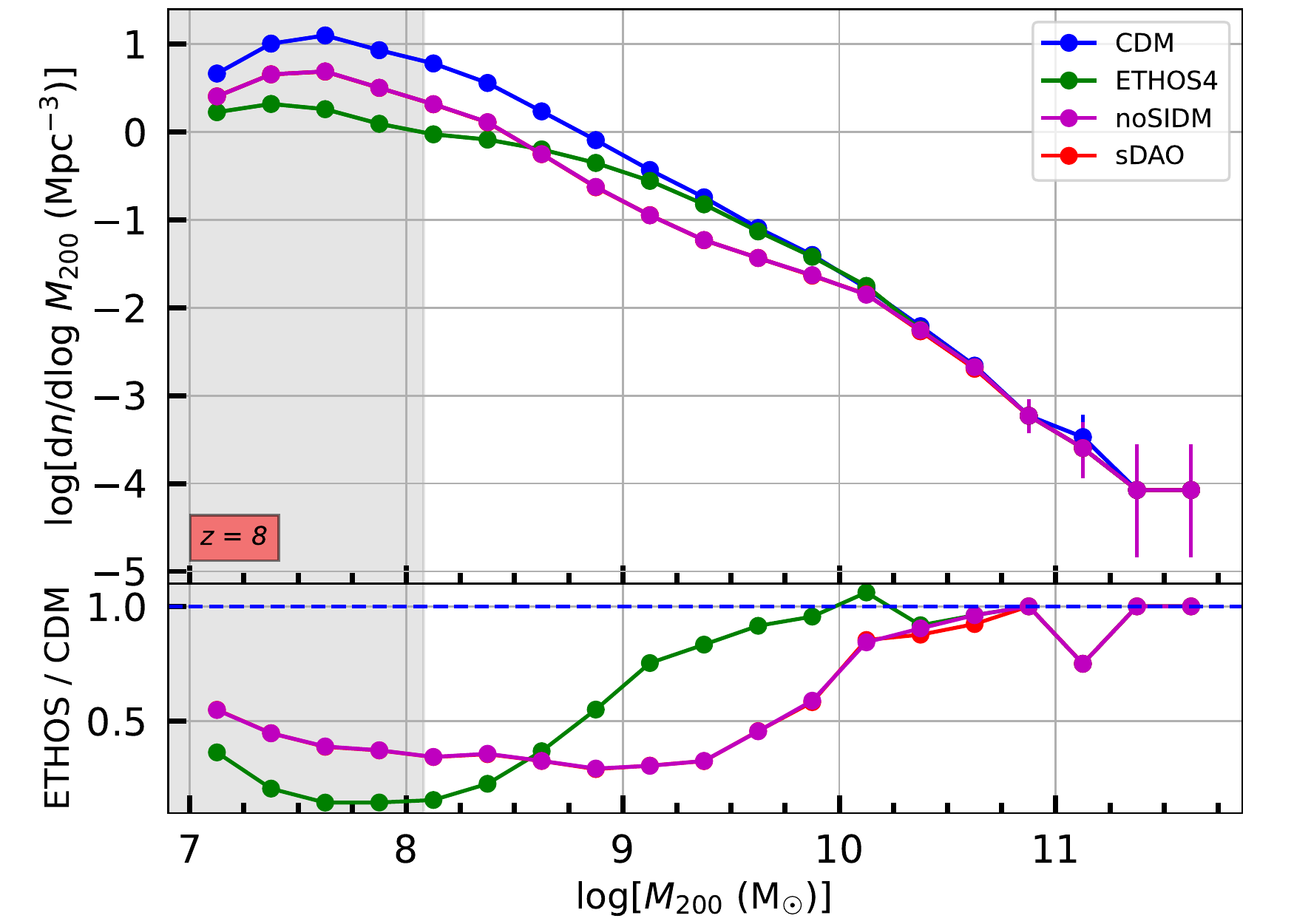}
    \\
    \includegraphics[width=\columnwidth]{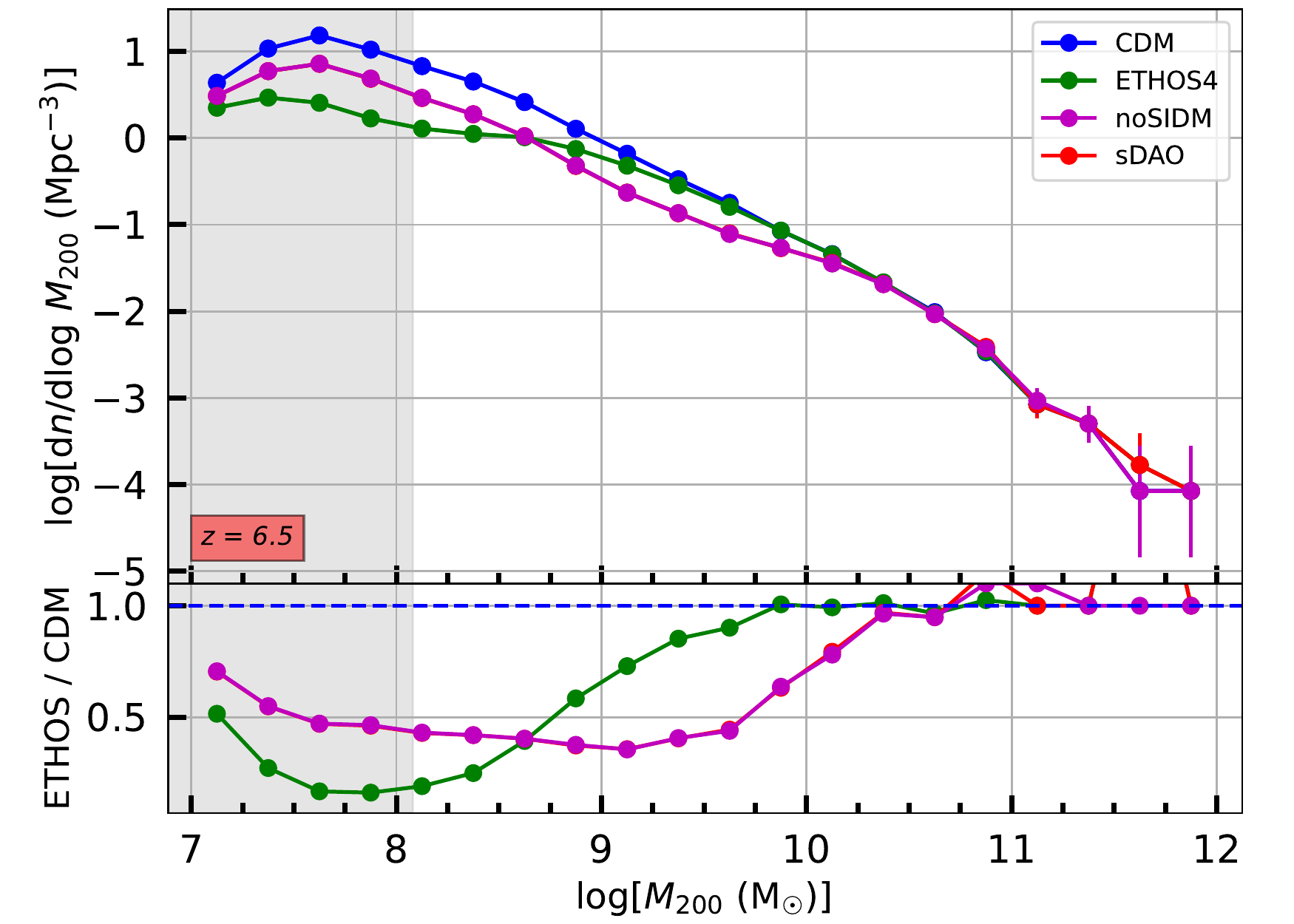}
    \caption{The DM halo population density in different mass ranges for redshifts 10, 8, and 6.5 for the different DM models we study (see Table~\ref{tab:dm_models}). The bottom panels show the ratio of particular models with respect to CDM. The horizontal axes are logarithmic mass ranges in units of solar masses. The vertical axes are the number of galaxies per unit volume per logarithmic bin. The error bars on the points depict the Poisson error in each mass bin. The grey shaded region indicates the regime where spurious fragmentation is expected to dominate the halo population in ETHOS4. The process for identifying this mass scale is described in the Section~\ref{sec:HMF}.}
    \label{fig:halo_mass_functions}
\end{figure}

Figure~\ref{fig:halo_mass_functions} portrays the abundance of haloes in different mass ranges. We  count the number of haloes in a particular logarithmic mass range. Then, to normalize, we divide the count by the simulation box size and the bin width. The so-called halo mass function is one of the primary statistics that can be used to compare the predictions of different DM models. We use $M_{200}$ to define halo mass, which is the mass of the halo at the radius within which the density is equal to 200 times the critical density of the universe.

Figure~\ref{fig:halo_mass_functions} shows that in mass ranges above $\sim10^{10}\,{\rm M}_\odot$ the mass functions for the four models are nearly identical, especially within Poisson errors marked using error bars in this figure. Despite the theoretically expected latency in structure formation in the ETHOS models, we do not find an obvious difference in the maximum halo mass formed at any particular redshift. This is not unexpected; Figure~\ref{fig:power_spectra} shows that the different power spectra are identical on large scales; therefore, one should not expect any differences in structure formation at the high mass end. In addition, \citet{Lovell_2019} explicitly shows that the latency is stronger for lower mass haloes than for higher mass haloes. Furthermore, the curves of alternative models become more like CDM at lower redshifts, indicating alternative models catch up with the delayed start of their structure formation by having a more rapid formation progress in the later redshifts. This phenomenon is also indicative of why the differences in the models would be most prominent at higher redshifts. Finally, the sDAO and noSIDM halo mass functions are identical. This suggests that at high redshifts, because the probability of DM collisions is low due to low DM density, effect of DM-DM self interactions could be considered negligible compared to the effect of DM-DR interactions.

 The characteristic mass scale where the differences between CDM and ETHOS models begin to appear is $\sim10^{10}\,{\rm M}_\odot$. From a physical perspective this is expected because one of the purposes of alternative models is to suppress dwarf scale structures. While ETHOS4 has a nearly identical HMF to CDM going to smaller scales compared to sDAO models, it has a sharper deviation starting from $\sim10^{8.5}\,{\rm M}_\odot$. This reflects the behaviour of the linear power spectrum in ETHOS4 which shows a sharp suppression of power after the peak going towards smaller scales. In contrast, the strong DAOs in the power spectrum of the sDAO model show a resurgence of power on small scales that pushes the sDAO model closer to the values for CDM. \citet{Bohr_2021} provides a detailed discussion of this power resurgence and shows that it is not a numerical artefact.  

On scales smaller than the grey shaded area, an upturn at low masses in the mass function of models with initially truncated power spectra is often the hallmark of numerical ``spurious'' fragmentation \citep[e.g.][]{Wang2007,Lovell2014}. To test whether the feature in Fig.~\ref{fig:halo_mass_functions} is  physical or numerical, we follow \cite{Wang2007} and compute the expected spurious fragmentation scale of our simulations:

\begin{equation}
\label{eq:resolution}
    M_{{\rm lim}} = 10.1\bar{\rho} d k_{{\rm peak}}^{-2}
\end{equation}
where $\bar{\rho}$ is the mean matter density of the universe, $d$ is the mean particle separation, and $k_{\rm {peak}}$ is the comoving wave number $k$ at which the dimensionless power spectra $\Delta^2(k)$ reaches its maximum value. The power spectra of all the models are shown in Fig.~\ref{fig:power_spectra}.

ETHOS4 has a clear peak before the DAOs start. On the other hand, sDAO in addition to the peak before the DAOs, has another peak that is higher than the initial peak within the DAOs. Thus we have calculated two different fragmentation scales for sDAO. The values are presented in Table \ref{tab:minimum_res}.
\begin{table}
    \centering
    \begin{tabular}{c|c}
        \hline
        Model & Min Resolution [${\rm M}_\odot$] \\
        \hline
        CDM & - \\
        ETHOS4 & $1.2 \times 10^8$ \\
        sDAO & $3.6 \times 10^8$ \\
        sDAO (Max) & $3.5 \times 10^6$\\
        \hline
    \end{tabular}
    \caption{The spurious fragmentation scales for different DM models. sDAO is calculated by taking the $k_{{\rm peak}}$ before the DAOs and sDAO Max was calculated with $k_{{\rm peak}}$ corresponding to the actual maximum.}
    \label{tab:minimum_res}
\end{table}
The values listed in Table~\ref{tab:minimum_res} suggest that the abundance of haloes below the crossing scale of the ETHOS4 and sDAO  curves could be spurious, and therefore do not reflect the true model predictions. However, this is not a problem for our present work because we are only interested in galaxies that will be observed by JWST, which will typically be hosted in haloes that are more massive than $\sim10^9 {\rm M}_\odot$. 

\subsection{Observational verification of galaxy populations}
\label{sec:Observational}

The primary objective of the present work is to compare the statistics of the galaxy populations in the different DM models of interest. To this end, each simulation has been run with a galaxy formation model very similar to that applied in the IllustrisTNG simulations. An important issue to consider is the extent to which the parameters of the model need to be calibrated for each individual DM model. Here, we have opted not to recalibrate, and instead use the best-fitting IllustrisTNG parameters for CDM in all our DM variants. This choice is justified by the fact that the regime where these models begin to deviate from CDM corresponds to galaxies that are much fainter than the set of galaxies used to calibrate the IllustrisTNG model. The parameters of the model are calibrated using data from bright galaxies i.e., in the regime where the different DM models are largely identical. Present high redshift observational data do not set stringent constraints in the regime of fainter galaxies where the models deviate more strongly; it would therefore be interesting to revisit the question of parameter recalibration when future data has more constraining power in the low mass regime.

Nevertheless, it is important to verify that this approach is valid, and that the galaxy populations in CDM and all ETHOS variants have properties that are consistent with existing data at the redshifts of interest. To this end, we compare the predictions of our simulations for the stellar mass functions and stellar-to-halo mass relations at $z=6-10$ against existing data.

\begin{figure}
    \centering
    {\includegraphics[width = \columnwidth]{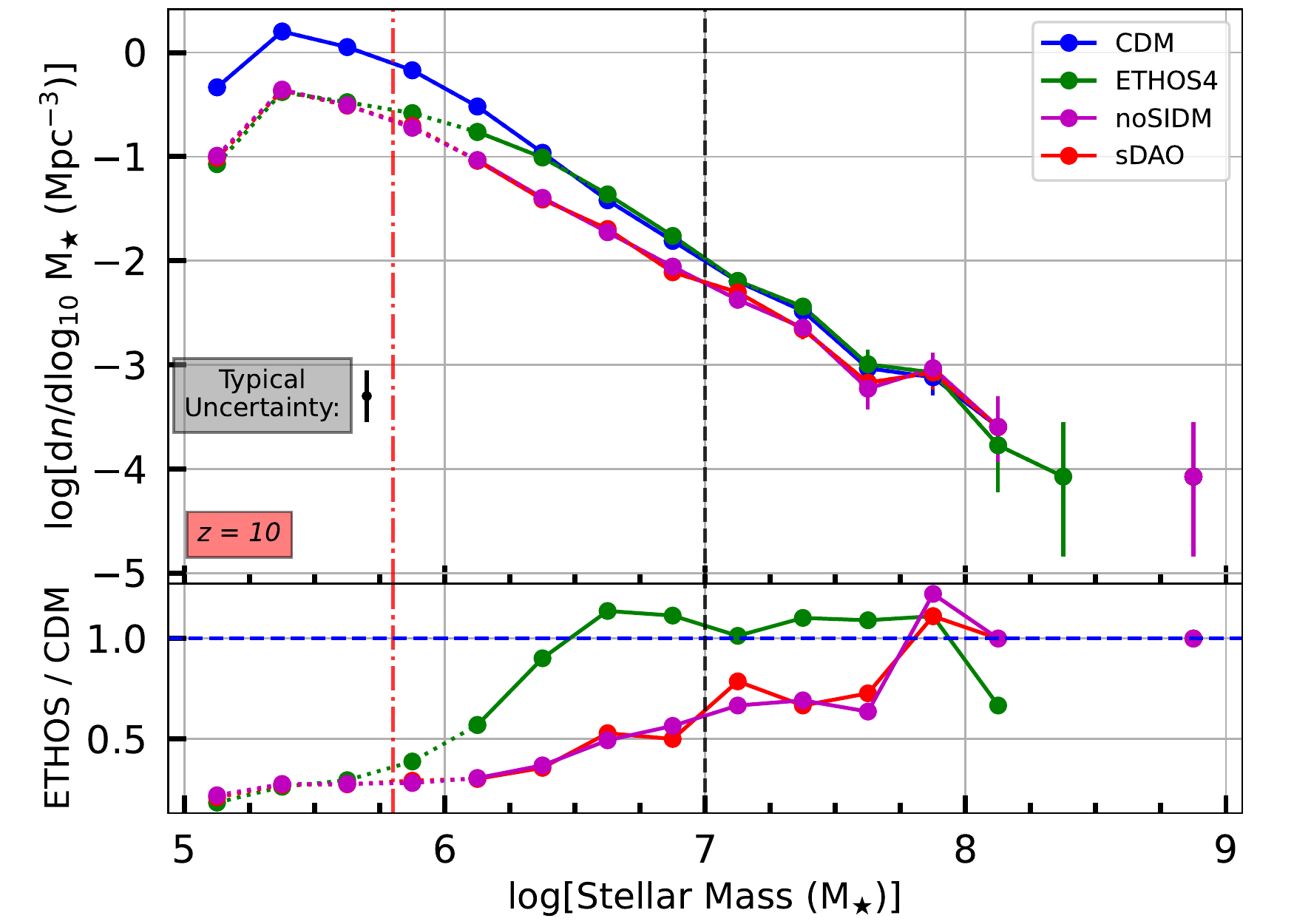}}
    \\
    {\includegraphics[width = \columnwidth]{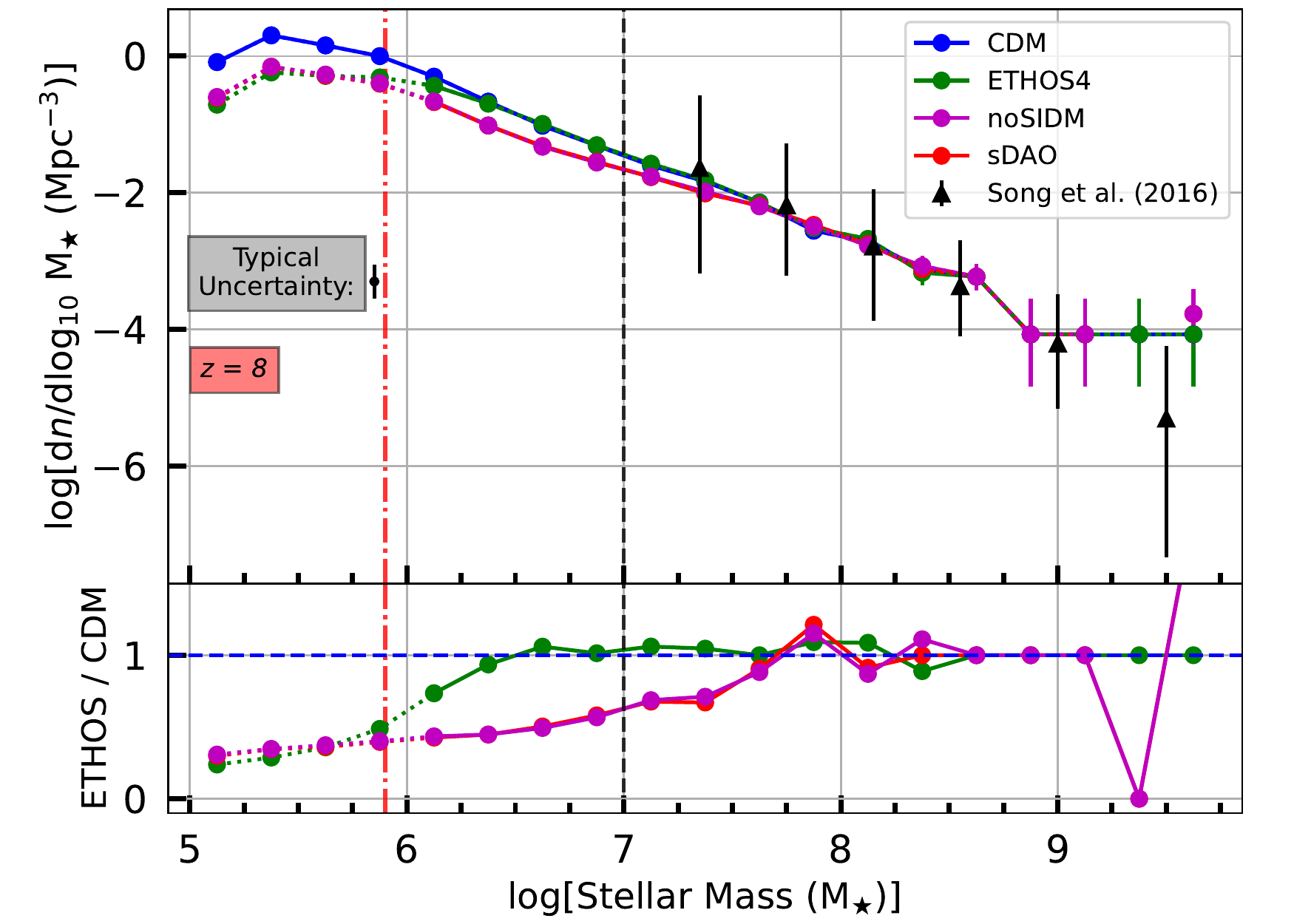}}
    \\
    {\includegraphics[width = \columnwidth]{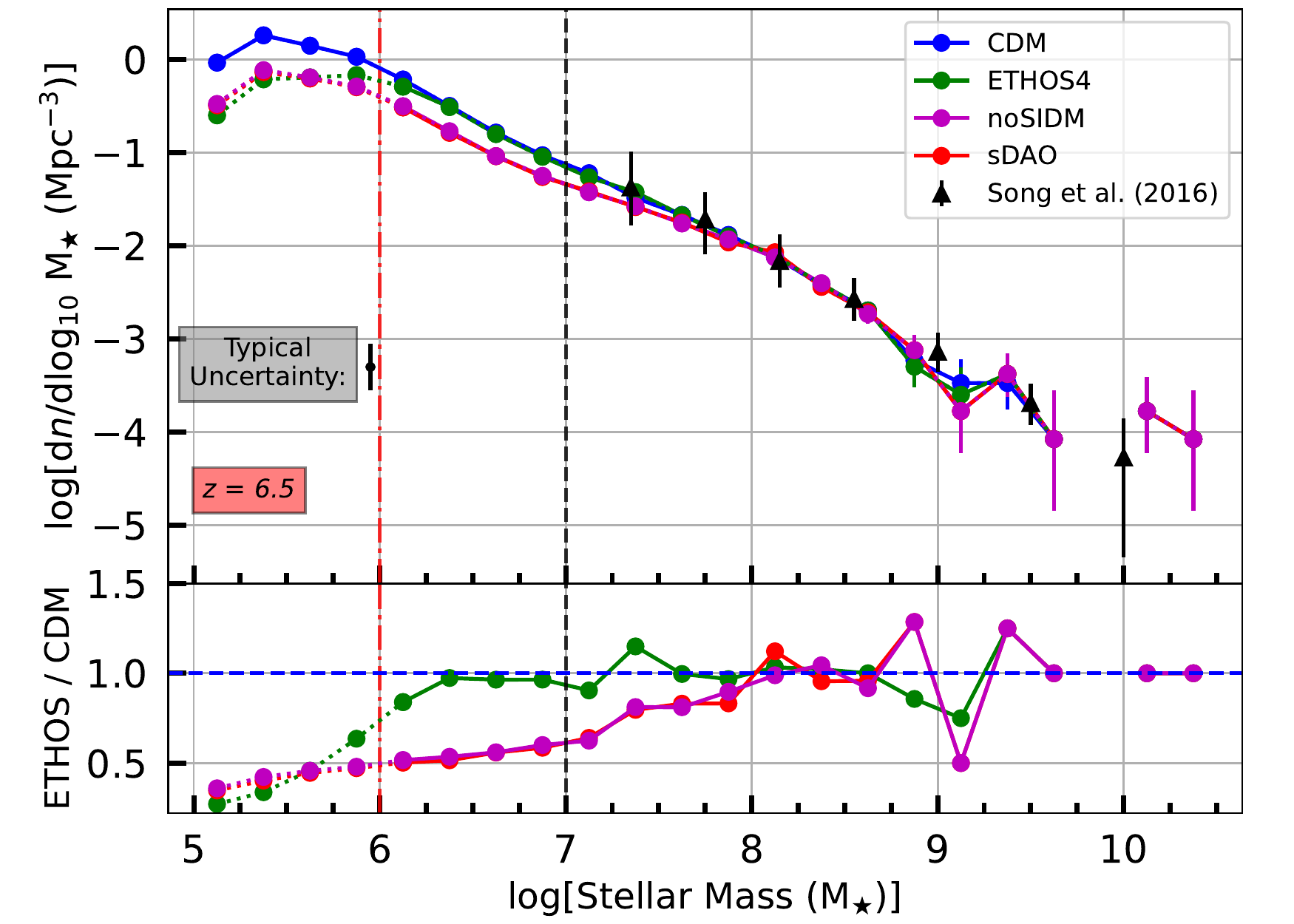}}
    \caption{The galaxy population density in different mass ranges for redshifts 10, 8, 6.5. The bottom panels show the ratio of particular models with respect to CDM. The horizontal axes are logarithmic mass ranges. The vertical axes are the number of galaxies per unit volume per logarithmic mass bin. The black and red dashed lines, respectively, represent the JWST sensitivity limit for an HUDF-like survey and a survey that includes strong gravitational lensing \citep[see, e.g.][]{Williams_2018,Yung_2018, Yung_2019}. The JWST uncertainty is the theoretical uncertainty in SMF due to the uncertainty in Schechter fit \citep{Yung_2019}. It is an estimate of the typical systematic uncertainty associated with JWST number counts. The dotted portions of the lines are the possible interval of spurious fragmentation. The black errorbars and points are measurements of \citet{Song_2016} that we use to verify our galaxy populations.}
    \label{fig:stellar_mass_functions}
\end{figure}

\begin{figure}
\centering
    {\includegraphics[width = \columnwidth]{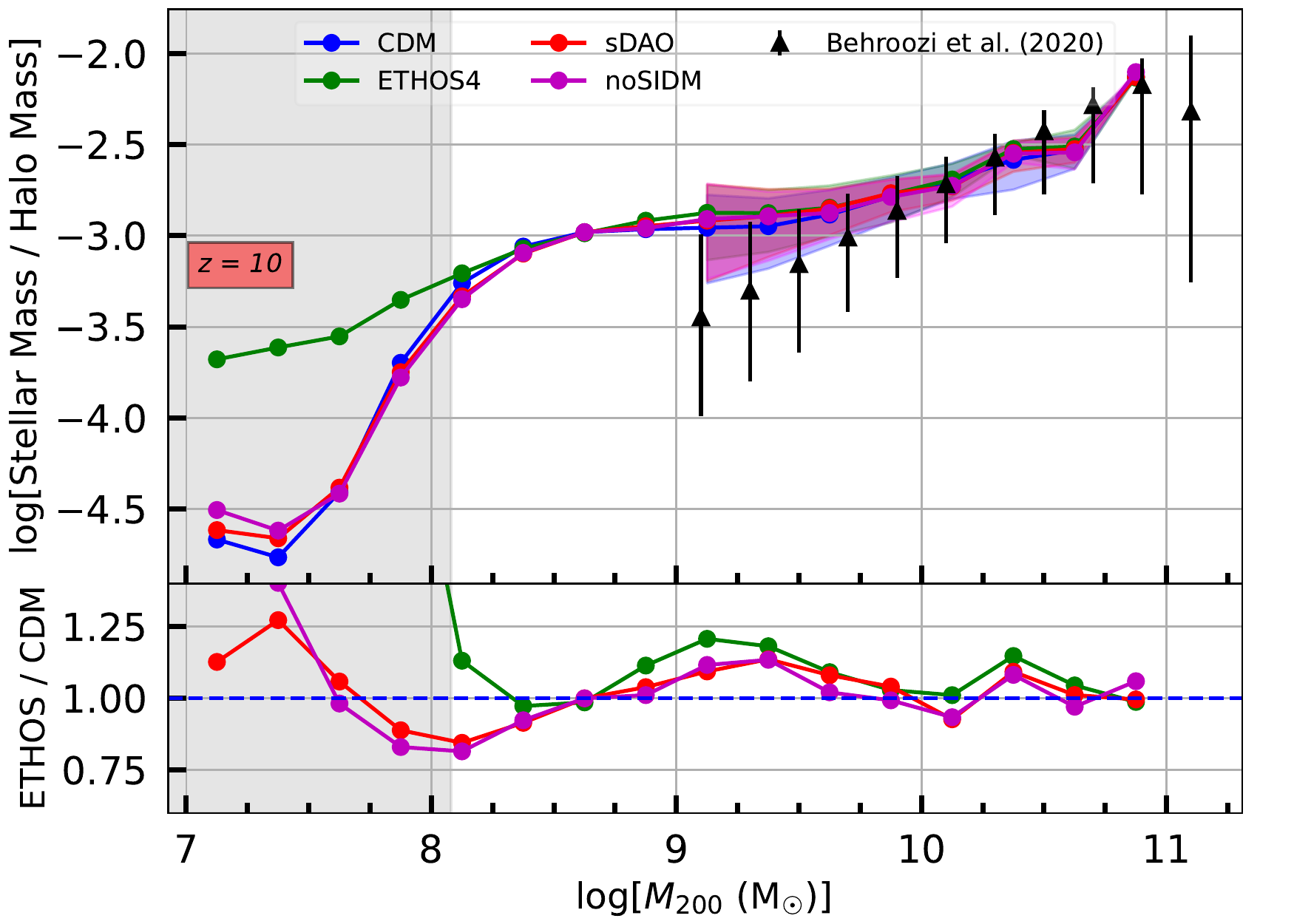}}
    \\
    {\includegraphics[width = \columnwidth]{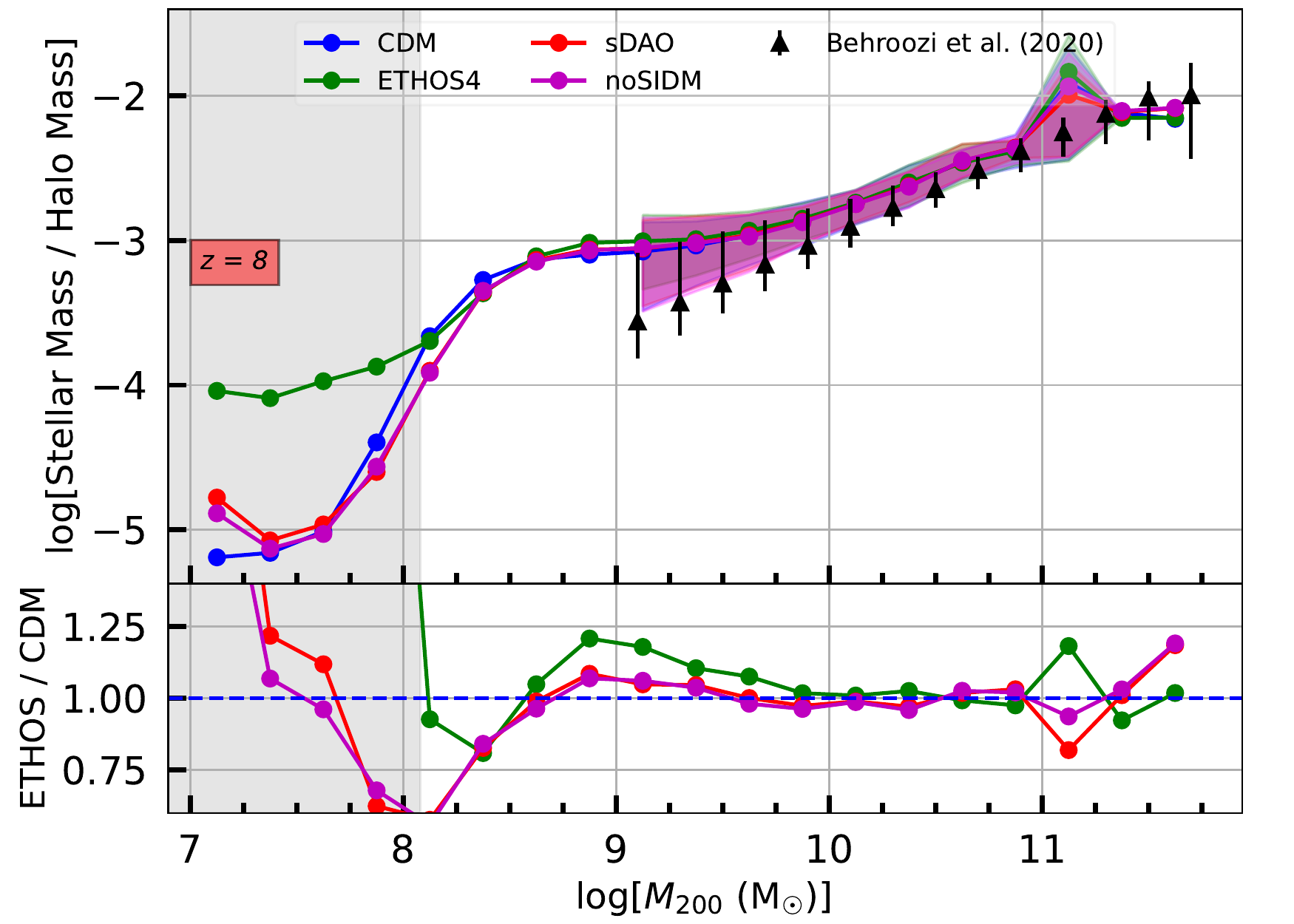}}
    \\
    {\includegraphics[width = \columnwidth]{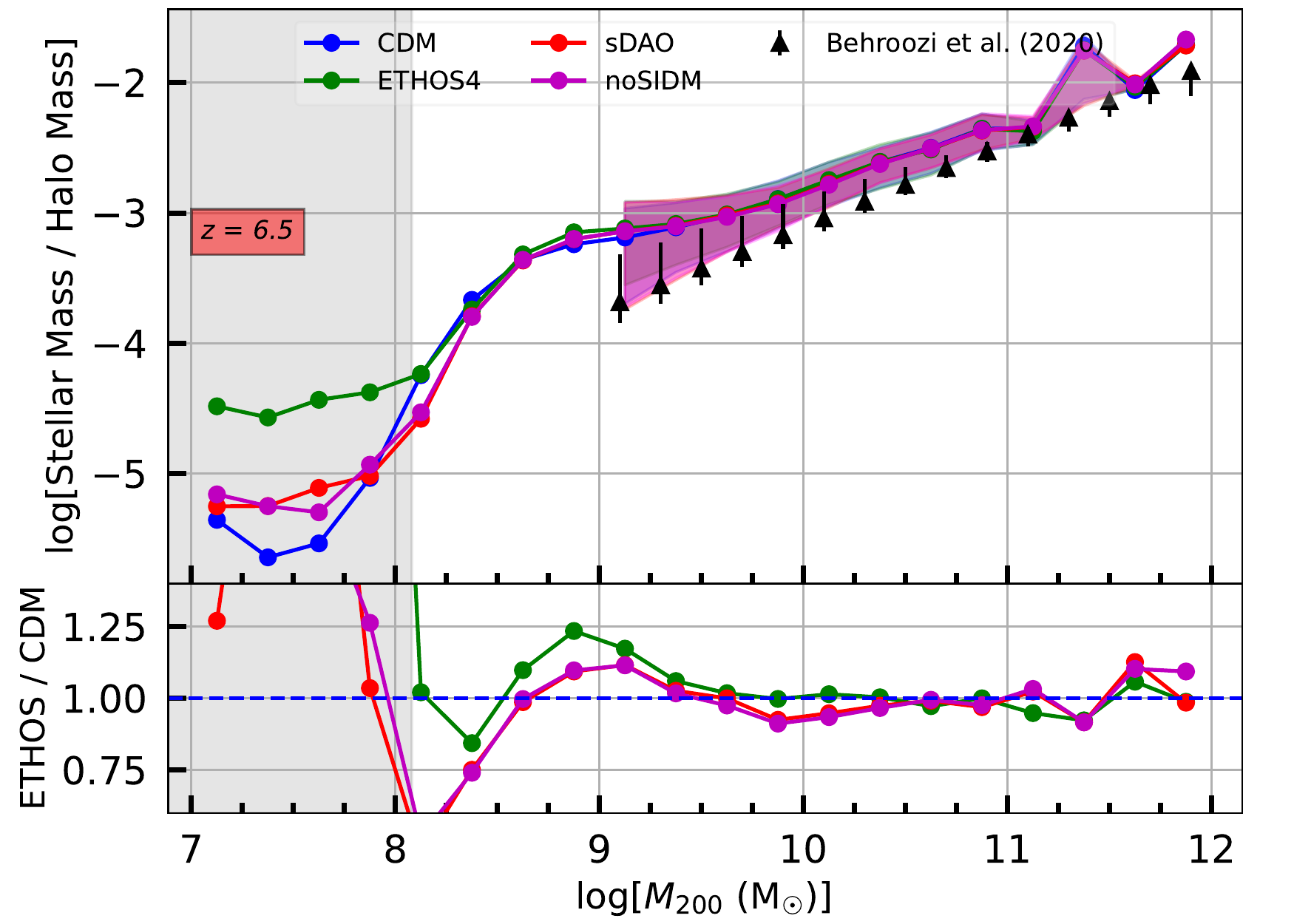}}
    \caption{The ratio of the galaxy mass to its host halo for redshifts 10, 8, and 6.5. horizontal axes are halo mass, $M_{200}$,  and the vertical axes are the ratio between stellar mass and halo mass. The top and bottom parts of the regions are the 84th and 16th percentile. The colored lines are the mean values of the ratios for that mass range. The bottom plot shows the ratio of ETHOS models to CDM. The gray shaded region is the interval of spurious fragmentation. The black points and errorbars are the measurements from \citet{Behroozi_2020} that we used for verification of stellar-to-halo mass comparisons.}
    \label{fig:halo_luminosity}
\end{figure}

Figure~\ref{fig:stellar_mass_functions} compares the stellar mass functions predicted by our models (coloured lines) to observational data from \citet{Song_2016} (black symbols). We find typically good agreement between the models and the observations in the regions where observational data are available. The models only begin to show significant deviation on smaller scales, which will be accessible by JWST. The dashed black and red lines on the panels represent the JWST sensitivity limit for an HUDF-like survey and measurements augmented by strong gravitational lensing \citep[see, e.g.][]{Williams_2018,Yung_2018, Yung_2019}. The error bars in the figure represent the theoretical expected scatter of stellar mass functions as computed by \citet{Yung_2019}. The dotted portions of the curves are the possible spurious fragmentation discussed in \ref{sec:HMF}. Because the simulation resolution limit is in terms of halo mass, and halo mass and galaxy mass do not have a constant relation, we used dotted lines in the possible region rather than completely shading it to portray that it is not a clear cut lower limit. To estimate the mass scale in terms of stellar mass, we use the stellar-to-halo mass relation (shown later in Fig.~\ref{fig:halo_luminosity}). Based on this, we conclude that the galaxy stellar masses are approximately 2.5 orders of magnitude lower than their host haloes' total masses.

On large scales, well above the JWST detection limit, the stellar mass functions for all the models are nearly identical especially within their individual Poisson uncertainties. This shows that although the DAOs in the initial power spectrum cause a suppression in the stellar mass functions, the scale of suppression is not sufficient for observationally differentiating underlying DM models with JWST for the brightest galaxies. On the other hand, for galaxies of stellar mass $M_\star \sim10^7-10^8 {\rm M}_\odot$ which will be detectable by JWST, the sDAO models (with and without SIDM) have around half as many galaxies as in CDM or ETHOS4. This is the case for all the redshifts shown in the panels.

Although a difference of a factor of 2 suggests JWST observations could be used to differentiate between the models, this is still smaller than the minimum $\pm 0.5$ dex theoretical uncertainty of JWST caused by the uncertainty of Schechter fit from \cite{Yung_2018}, which is driven by theoretical uncertainties in modelling faint galaxies. On the other hand, between galaxies with stellar mass $M_\star \sim10^6-10^7 {\rm M}_\odot$--between the strongly lensed and non-lensed limits--the difference between CDM and extreme models are consistently greater than the uncertainty. This shows that with the help of strong gravitational lensing, the faint end of the stellar mass functions may indeed provide a feasible way to constrain dark matter models using JWST. That being said, it is important to bear in mind that such a difference is a result of a model designed to create prominent galactic scale differences. The sDAO models can therefore be considered as a plausible upper limit of the impact of the DM coupling with relativistic particles in the early Universe. In Section~\ref{sec:Dark_Matter} we showed that the interaction rates of such coupling in sDAO models are 1 million times greater during the epoch of decoupling than the fine-tuned ETHOS4 models. A less extreme model may therefore be indistinguishable, at least in terms of galaxy abundance.

In addition to their absolute normalization, the relative shapes of the curves in Fig.~\ref{fig:stellar_mass_functions} also provide useful information. We see that between the JWST resolution limit and the numerical fragmentation limit, the slopes of the sDAO curves increase while slopes of the CDM and ETHOS4 curves are constant above the simulation resolution limit. This is the result of the power peeks at small scales caused by DAOs.

For our next comparison of galaxy properties in these DM models, we present the stellar-to-halo mass relations, depicted in Fig.~\ref{fig:halo_luminosity}. To account for the halo mass, we again use $M_{200}$. For the galaxy masses, we take the stellar mass of the most massive subhalo within a given halo. As a result, we only consider the relation of central galaxies and their host haloes, which is the subset of the galaxy population that dominates the stellar mass function. %Despite the exclusion of satellite galaxies, this section still accurately evaluates the relation between the Stellar Mass and Halo Mass functions since the Stellar Mass Function is mostly composed of central galaxies.

Figure~\ref{fig:halo_luminosity} shows that the stellar-to-halo mass relations are almost completely insensitive to the underlying particle physics models. At each redshift, all the models follow a similar trend and, in fact, they almost have identical values. There is a steep drop in stellar to halo mass ratio below $M_\star\sim10^{8.5}\,{\rm M_\odot}$. This is due to haloes below that mass very rarely hosting galaxies and their mean being dominated by non-luminous haloes. Furthermore, we do not plot the scatter below this mass threshold because both the 16th and 84th percentile are 0. On the other hand, almost every haloe above $M_\star\sim10^{8.75}\,{\rm M_\odot}$ has a luminous component resulting in their data not being dominated by 0 entries. This contrast is quantitatively further discussed in Section~\ref{sec:occupancy}. Overall, not only the mean values are close, but scatter at fixed mass (indicated by the shaded bands) for each model is also very similar. The only exception to this is ETHOS4 having higher values below $M_\star\sim10^8\,{\rm M_\odot}$. We mention a plausible cause for this phenomenon in the discussion of Fig.~\ref{fig:sfr_distribution}, but do not further discuss it because it is in the scale affected by finite simulation resolution and limitations of our galaxy formation model. This suggest that even though the abundance of haloes with respect to their mass were affected by particle physics, the internal properties of haloes such as the stellar-to-halo mass relation is likely more strongly affected by the galaxy formation physics.

%Along with the further insight of macro scale similarity of the models, the 
Figure~\ref{fig:halo_luminosity} also %reinforces the formation of higher mass haloes over time and 
demonstrates that galaxies are generally at least two order of magnitude less massive than the haloes they are hosted in. This sets a nominal limit for a ``resolved'' galaxy in the simulation at approximately 10 star particles ($M_\star\sim10^6\,{\rm M_\odot}$), which would be approximately an order of magnitude lower than the observational limit of the JWST in our numerical setup.

Finally, the black symbols in Fig.~\ref{fig:halo_luminosity} show constraints on the stellar-to-halo mass relation at these redshifts as inferred from the {\it Universe Machine} project by \citet{Behroozi_2020}. Again, we find good qualitative agreement between these data and our simulation predictions, further justifying our choice not to recalibrate the IllustrisTNG model for each DM theory. The  \cite{Behroozi_2020} data and our models agree at the 1-$\sigma$ level.
\begin{figure}
\centering
    {\includegraphics[width = \columnwidth]{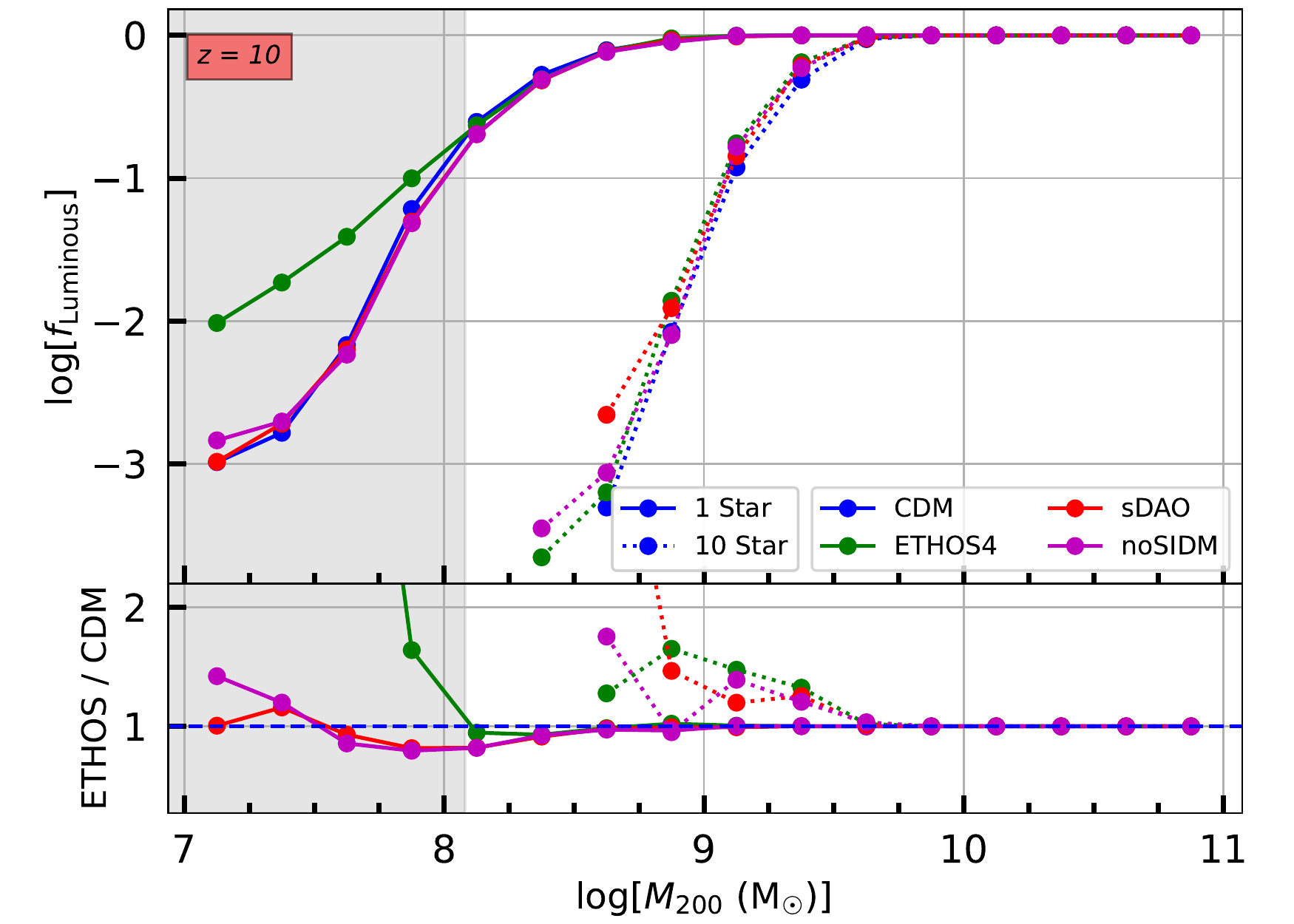}}
    \\
    {\includegraphics[width = \columnwidth]{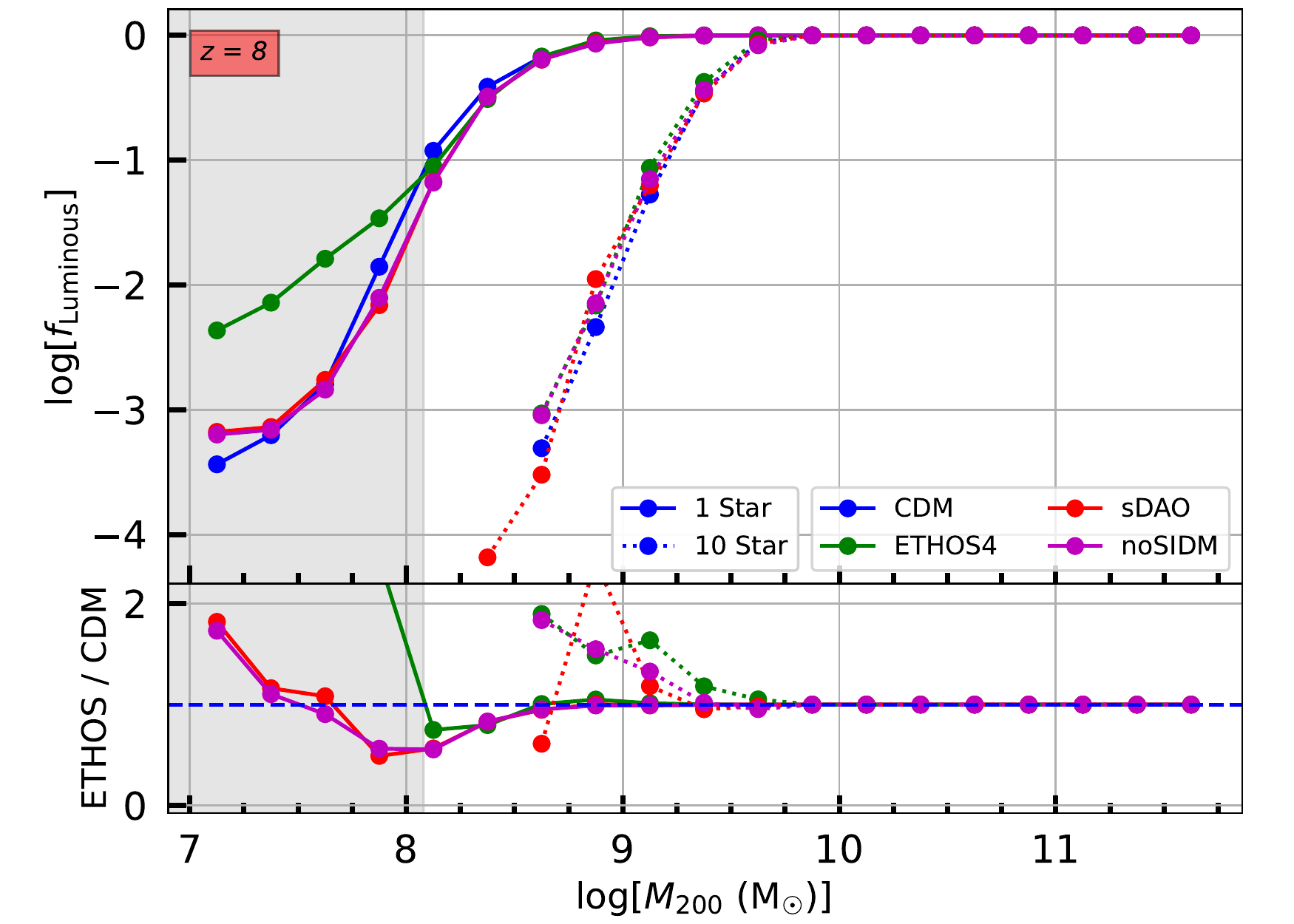}}
    \\
    {\includegraphics[width = \columnwidth]{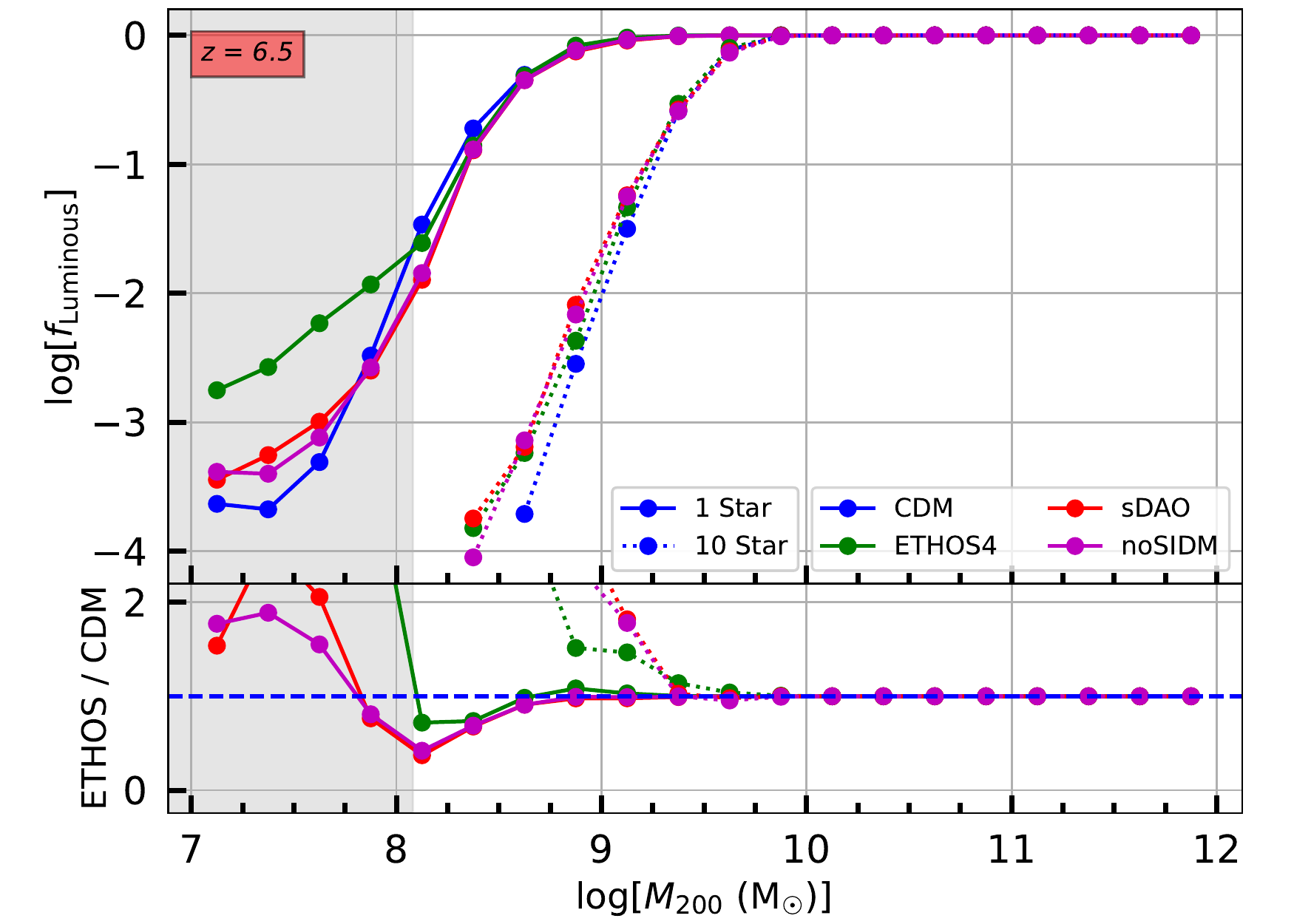}}
    \caption{The ratio of the number of haloes with at least 1 star particle (stellar mass $\sim 10^5\,{\rm M}_\odot$) to the total number of haloes in a particular logarithmic mass range. The dotted lines portray the same information about 10 star particle (stellar mass $\sim 10^6\,{\rm M}_\odot$) haloes. The three rows represent redshifts 10, 8, and 6.5. The gray shaded region is the interval of spurious fragmentation. We see that above $\sim 10^{10}\,{\rm M}_\odot$, all haloes host a galaxy which means there can be a one to one mapping from haloes to galaxies; this fraction drops sharply below $10^9\,{\rm M}_\odot$.}
    \label{fig:halo_star_ratio}
\end{figure}

\subsection{Galaxy-halo occupancy}
\label{sec:occupancy}

In the previous subsection, we have seen that despite the strong differences between the models in terms of their particle physics origins, the amount of DM-DR and DM-DM interactions do little to change how many stars form within a given halo. However, this neglects the fact that there could be a difference in the amount of haloes that {\it actually host} a galaxy. Therefore, we next consider a measure of the fraction of luminous haloes as a function of halo mass; this is shown in Fig.~\ref{fig:halo_star_ratio}.

The horizontal axes again represent the logarithmic halo mass range. The vertical axes represent the ratio of the number of haloes that host a galaxy to the total number of haloes within that particular mass range, denoted by the label $f_{\rm Luminous}$. To determine whether or not a halo hosts a galaxy, we investigate the presence (or not) of star particles in its central subhalo.

In cosmological simulations, there is inevitably some ambiguity when it comes to determining what (i.e. how many star particles) constitutes ``a galaxy''. Here, we consider two thresholds: one with the simplest case of a single star particle which roughly gives a sense of whether the minimal conditions for star formation are met within a given halo, and a more conservative choice of 10 star particles which corresponds to $10^6 M_{\odot}$. The former just considers an existence of a nonzero luminous component while the latter approaches to a more coherent definition of a galaxy. %It is also around the scale down to which we expect the IllustrisTNG model to converge at this resolution.

The primary insight from Fig.~\ref{fig:halo_star_ratio}, which shows the fraction of haloes of a given mass that are luminous (i.e. host a galaxy) is that for all models, high mass haloes (i.e., those that are above $10^{10}\, {\rm M}_\odot$) are nearly all luminous using both threshold definitions (1 or 10 star particles). However, the ability to host galaxies rapidly drops from 100 to 10 then to 1 per cent  within an order of magnitude of this mass scale. In addition, galaxy occupancy fraction in haloes does not evolve strongly as a function of redshift. Furthermore, we find that the behaviour is quite similar in all models considered here, irrespective of whether or not DAOs or self-interactions are present.

In conclusion, Figs.~\ref{fig:stellar_mass_functions}-\ref{fig:halo_star_ratio} show that, even for models that are made to create prominent galactic scale differences, the number of galaxies with respect to mass, galaxy-halo mass relation, and number of haloes that host a galaxy predicted by different DM models may be challenging to constrain observationally, even with JWST's capabilities. Nevertheless, the similarity in the stellar-to-halo mass relations and the galaxy-halo occupancy fractions means that it may be possible to infer the theoretical halo mass function in a way that is agnostic to variations between DM models.

\subsection{Breaking down the cosmic star formation rate (SFR)}
\label{sec:sfr_sum}

\begin{figure}
    \centering
    \includegraphics[width = \columnwidth]{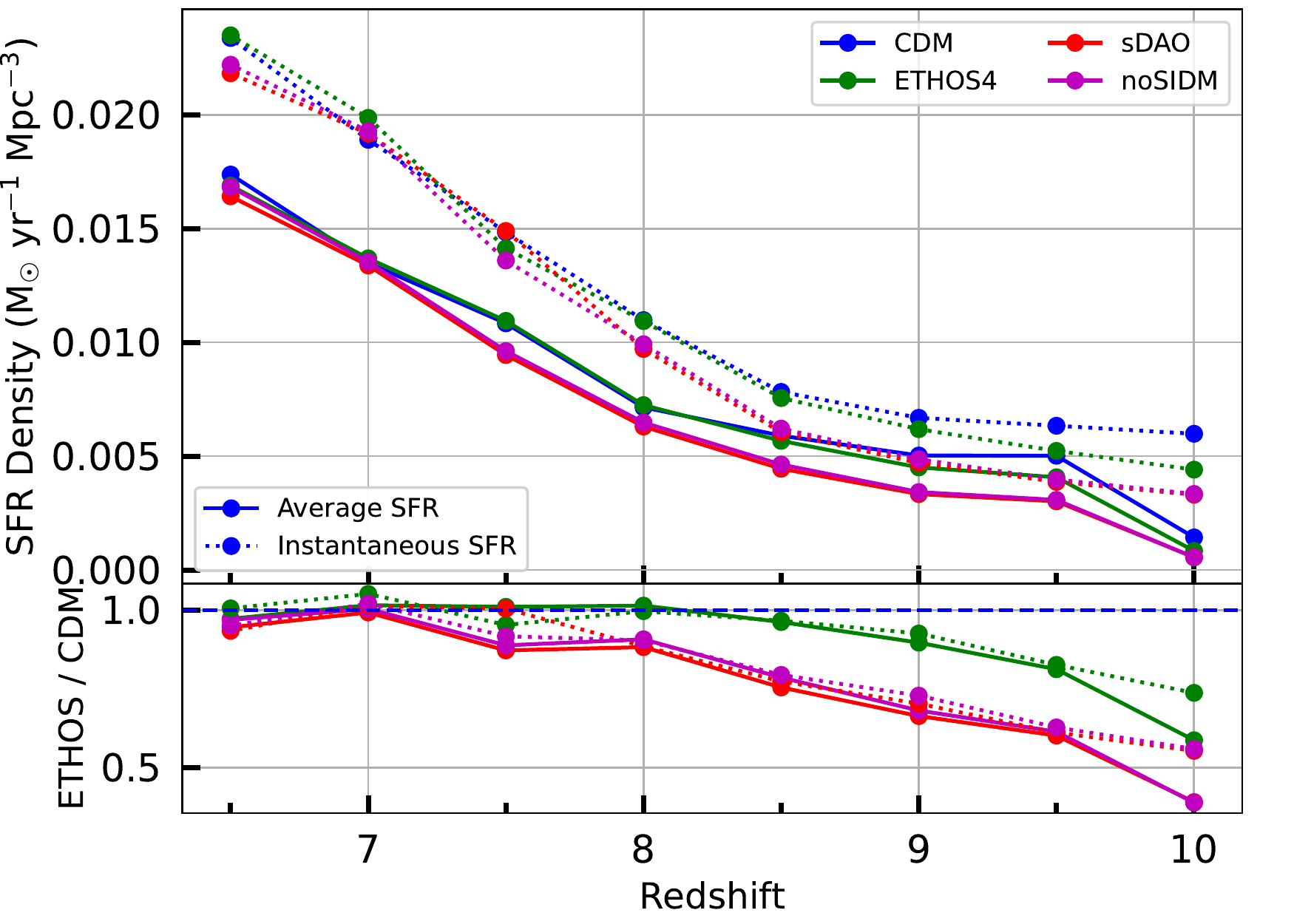}
    \caption{The total star formation rate of all the galaxies at a given redshift per unit volume (${\rm M}_\odot {\rm yr}^{-1} {\rm Mpc}^{-3}$). The dotted lines are the instantaneous SFR and the solid lines are the average SFR. The bottom panel shows the ratio of the ETHOS values to those in CDM. The average SFR was calculated by dividing the total stellar mass difference by the cosmic time difference in years. In this figure we see that the alternative models catch up to CDM as time progresses, despite the initial delay in the onset of star formation.}
    \label{fig:sfr_densityl}
\end{figure}

In all the previous figures, we have seen ETHOS models catching up to CDM in terms of the amount of structures with the progression of time. %This means there had been a difference in structure formation around those times.  JWST could measure the total star formation rate of a given snapshot. 
Investigating the history of cosmic star formation therefore provides a useful characterisation of the buildup of the galaxy population in each of the models considered here. This is measured as the star formation density (Fig.~\ref{fig:sfr_densityl}), which shows the sum of star formation rate per volume of all the galaxies in the simulation. The dotted lines are the {\it instantaneous} star formation rates associated with subhaloes in a given snapshot, while the solid lines are the average rates calculated by dividing the difference of total stellar mass of all the galaxies by the difference of cosmic lookback time (in years) in successive redshifts.

\begin{figure*}
    \centering
    \includegraphics[width = 0.9\textwidth]{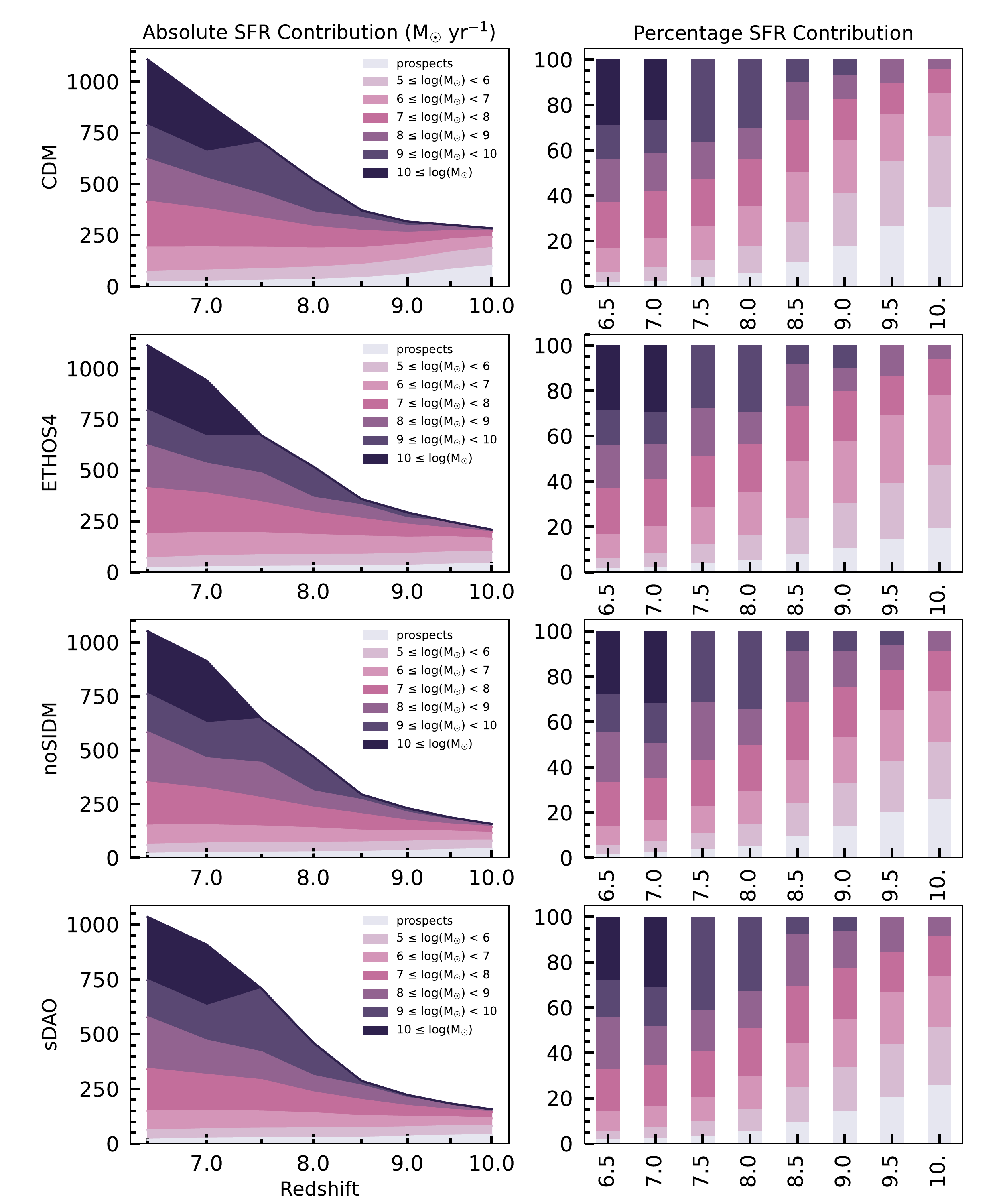}
    \caption{The contribution of galaxies in different mass ranges to the total star formation rate (${\rm M}_\odot {\rm yr}^{-1}$) as a function of redshift. The area plots on the left are the absolute contributions of different mass ranges to the total at that epoch. The ``prospect'' mass range consist of subhaloes that have just begun forming stars, but do not have a stellar mass yet. %The x axes are the redshifts and the y axes are the total star formation rate of all the galaxies. 
    The stacked bar plots on the right portray the percentage contribution of different mass ranges at different redshifts. Colours follow the same scheme as the panels on the left.}
    \label{fig:total_sfr_breakdown}
\end{figure*}
The total star formation rate always increases over time in all of the models indicating that universe is still in the process of building up stellar mass at these epochs. The instantaneous values always being larger than the average values also reinforces the constantly increasing stellar mass buildup for all the models. The effect of the strength of DM-DR coupling is very clear with Fig.~\ref{fig:sfr_densityl}. While the ETHOS4 catches up with CDM around redshift 8.5, sDAO catches up at  redshift 7. The DM-DR interactions clearly affect star formation rates. On the other hand, DM-DM interactions alone seem to have no effect on star formation rates, as becomes evident by the nearly identical behaviour of the sDAO and noSIDM models across all redshifts.

In addition to the absolute total, it is interesting to consider how individual galaxies contribute to the total star formation budget in each of the DM models we consider. This is especially pertinent considering the fact that while faint galaxies are expected to dominate by numbers (and therefore perhaps also in the total ionizing flux at these epochs) in a CDM universe, the abundance of these galaxies is largely suppressed in the ETHOS4 and sDAO models. In Fig.~\ref{fig:total_sfr_breakdown}, we show the total star formation density in our simulations, broken down into contributions from galaxies in different stellar mass ranges.

Figure~\ref{fig:total_sfr_breakdown} demonstrates that higher mass objects contribute more to the total star formation despite being less numerous. However, these high mass objects (galaxies above $10^9{\rm M}_\odot$) do not consistently exist until redshift 8. Thus the majority of the contribution at higher redshifts come from galaxies that are below $10^7\,{\rm M}_\odot$. Figure~\ref{fig:stellar_mass_functions} shows that CDM has twice as many galaxies in that mass range. This directly results in CDM having a total star formation rate density twice as large as the other models at earlier redshifts. This means total ionization flux in a CDM universe above redshift 8.5 is larger than any (including the fine-tuned ETHOS4) model that includes DAOs in the initial power spectrum. Furthermore, the difference can go up to a factor of 2 with models that have strong dark acoustic oscillations as seen with the sDAO models in Fig.~\ref{fig:sfr_densityl}. This suggests that, within the model used, CDM universe is going through reionization earlier than in the ETHOS models considered here. 

We have mentioned that at higher redshifts the majority of star formation contribution came from low mass galaxies; however the fraction of the total budget contributed by these low mass galaxies varies between models. In particular, while galaxies less massive than $10^6\, {\rm M}_\odot$ make up more than 60 per cent of the total star formation at redshift 10 in CDM, the contribution of that mass range is below 50 per cent for all the other alternative models. Conversely, both the absolute values and the mass range distributions of all the models are extremely similar at lower redshifts. Despite the vast differences in the primordial/linear power spectrum across DM models, the detailed breakdowns of the total star formation rate do not differ significantly. 

Although the summed contribution of particular mass ranges being similar in different models indicates a similarity between the models, from a statistical point of view, the summed totals of two data sets are not the optimal way to compare them. Figures~\ref{fig:stellar_mass_functions} and~\ref{fig:total_sfr_breakdown} show that models exhibit similar amounts of total star formation from similar amounts of galaxies, yet they do not provide information on the detailed distribution of star formation within a mass range. For example, it is possible that DM-DR or DM-DM interactions may affect the scatter of star formation rate within a mass range. To better understand the variation of star formation among the models, we next consider the distributions of star formation rates for different mass ranges.

\begin{figure*}
	\includegraphics[width=\textwidth]{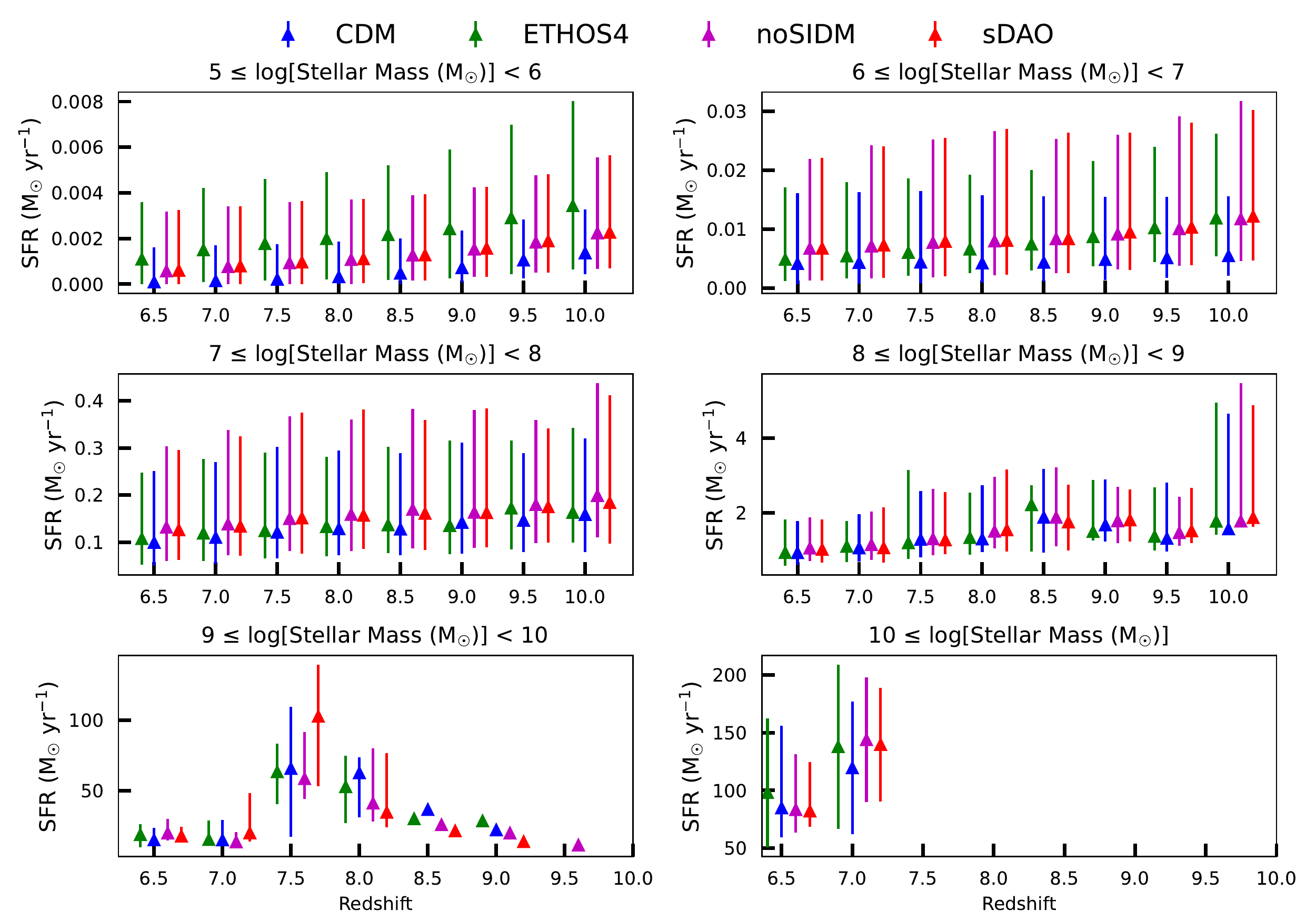}
    \caption{The subfigures show the distributions of star formation rates of galaxies in a particular mass range and at a particular redshift for different DM models. The middle of the errorbar is the median, the top is the 84th percentile, and the bottom is the 16th percentile. The vertical axes are the star formation rates of individual haloes with the dimensions of solar mass per year (${\rm M}_\odot {\rm yr}^{-1}$). As the mass of the galaxies increase in order of magnitude, the star formation rates increase in order of magnitude.}
    \label{fig:sfr_distribution}
\end{figure*}

Figure~\ref{fig:sfr_distribution} represents the star formation rate distribution of individual subhaloes with respect to redshift and mass range. It provides a comparison of the properties of galaxies in a narrow mass range between the models. The distribution is represented as error bars and points, where the top of the error bar corresponds to the 84th percentile and the bottom corresponds to 16th percentile of the data while the actual data point corresponds to the median value.

 Figure~\ref{fig:sfr_distribution} clearly demonstrates the positive correlation between the star formation rate and galaxy mass. As the galaxy mass increases by an order of magnitude the star formation rate also increases by an order of magnitude. The first three subfigures have very large errorbars because the stochastic scatter in SFR rates is much larger than the systematic difference in the median SFRs of the various model / redshift combinations. In the second and third figures (mass range from $10^6$ to $10^8M_{\rm \odot}$), although a very minor difference, the median and the 84th percentile SFR value increases with the increased deviation of the model from CDM. sDAO has the highest values for both. The distributions do not show any significant differences in the rest of the plots for the higher masses. Again we see very similar distributions despite the differences in the linear power spectrum, just as we have observed in our previous results. 
 
 The only major difference is sDAO having a markedly increased SFR in the mass range $10^9-10^{10}\,{\rm M_\odot}$ at redshift 7.5. One may think that this outlier is the reason that made sDAO structurally catch with other models in terms of structure formation. However, the rapid growth of the star formation rate between $z = 9$ and $z = 7$ in Fig.~\ref{fig:sfr_densityl} occurs for both sDAO and noSIDM, whereas the spike for $10^9$ to $10^{10}M_{\rm \odot}$ at $z=7.5$ occurs only for sDAO. A much more likely reason is the consistent offset of $\sim 0.02M_{\rm \odot} {\rm yr}^{-1}$ between sDAO/noSIDM on the one hand and CDM/ETHOS4 on the other in the $10^7$ to $10^{8}M_{\rm \odot}$ range. Even though the sDAO/noSIDM median lies within the distribution of ETHOS4/CDM, the distribution as a whole is shifted to higher star formation rates and this causes the accelerated cosmic star formation rate. At lower masses, SFR in the ETHOS4 model is significantly higher than CDM too, which drives the ETHOS4 model to catch up with CDM at $z=8.5$.
 
In the mass range $10^5-10^6\,{\rm M_\odot}$, ETHOS4 has consistently higher star formation rate at all the redshifts. The SFR of galaxies are strongly determined by the amount of cold gas available. This suggests that ETHOS4 haloes have more gas available than their CDM counterparts despite having less DM mass. \citet{Lovell_2018, Lovell_2019} discusses this phenomenon of increased availability of gas (and resultant star formation through starbursts) during merging processes in detail. Although it has a physical explanation, this phenomenon occurs in mass ranges where our galaxy formation model is not fully trustworthy.

%\newpage
\subsection{The clustering of galaxies at high redshift}
\label{sec:clustering}

Historically, the distributions of galaxies have been used to constrain cosmological models because the degree of clustering is sensitive to the underlying cosmology. However, clustering measurements at higher redshifts have been limited in scope due to the finite availability of resolved galaxies. With JWST, the potential of these measurements will be significantly enhanced. 
\citet{Endsley_2020, Vogelsberger_2020} provide a quantitative demonstration of the expected improvement and \citet{Sabti_2021} suggests an additional method to increase the sensitivity of these measurements at high redshifts.

The {\it Hubble Space Telescope} (HST) has a primary mirror with a diameter of 2.4 meters. The mirror of JWST has a diameter of 6.5 meters which will approximately give it 6.25 times more light collecting area. Furthermore, JWST is equipped with an NIRCam that can take photographs of the earliest galaxies. Quantitatively, deep (m $\gtrsim$ 29-30) NIRCam photometry from Cycle 1 programs will enable the detection of  $\sim$ 5000 galaxies at $z>6$ and $\sim$ 300 at $z>9$ \citep{Williams_2018}. These increased numbers will highly benefit clustering measurements because Poisson noise decreases linearly with the inverse of galaxy density \citep{Peebles_1980, Landsy_1993}. In addition, the clustering calculations of lower mass galaxies will be less affected by cosmic variance \citep{Sommerville_2004, Trenti_2008}. Furthermore, JWST will enable the first spatial galaxy clustering measurements at $z>4$ with its multiplexing capabilities and sensitivity to strong rest-optical lines at these high redshifts \citep{Chevallard_2019, DeBarros_2019}. Spatial clustering measurements will be able to avoid chance projections and low redshift contaminants that are present in angular clustering measurements. Consequently, it would be possible to measure clustering to $\sim 4-5 \sigma$ significance up to $z\sim 10$. This is initially argued by \citet{Zhang_2019} based on cosmological N-body simulation results. \citet{Endsley_2020} further reinforces it by adding a model of high redshift galaxy selecting, simulating clustering measurements on exact footprint of JADES program, and including satellite galaxies in the simulated measurements.

Galaxy clustering measurements may be used to infer halo masses of the galaxy samples in question by exploiting the strong relation between halo mass and clustering strength \citep[e.g.][]{MoWhite_1996, Tinker_2010}. \citet{Wechsler_2018} provides a review of this connection. Here, we present a first attempt to quantify the clustering strength of galaxies in exotic DM scenarios at higher redshifts. A difference in clustering statistics between the models could be a novel way to constrain the nature of DM in the era of galaxy redshift surveys.

We compute the two-point clustering statistics using the publicly-available {\tt corrfunc} library \citep{Corrfunc_2020}. The resulting correlation functions measured for galaxies in different mass ranges are shown in Fig.~\ref{fig:raw_clustering}. In the panels of this figure, the different colours represent different models, and the shading of the lines represent the different threshold masses used to define the clustering calculation. Table \ref{tab:number_of_galaxies} presents the number of galaxies that was included in each combination of model, redshift, and low mass threshold plotted in the figure. The table also includes the total number of galaxies with a luminous component for a particular model and redshift. The table and figure together show the increasing unreliability of clustering calculations with decreasing number of elements included in the calculation.
\begin{table}
\centering
\caption{Number of galaxies included in the clustering calculation for a particular model, low mass threshold, and redshift. The leftmost column indicates the model and the redshift. The successive columns represent the low mass threshold applied in terms of stellar mass. The number of galaxies consistently decreases an order of magnitude with increasing lower mass threshold.}
\label{tab:number_of_galaxies}
\begin{tabular}{lcccc}
\hline
Model-Redshift & $10^0\, {\rm M}_\odot$ & $10^6\, {\rm M}_\odot$ & $10^7\, {\rm M}_\odot$ & $10^8\, {\rm M}_\odot$ \\
\hline
CDM-10& 51148 & 5660 & 138 & 4 \\
ETHOS4-10 & 17089 & 4076 & 145 & 4 \\
noSIDM-10 & 14379 & 2000 & 99 & 4 \\
sDAO-10 & 14281 & 1982 & 107 & 4\\
CDM-8 & 73320 & 10832 & 631 & 43 \\
ETHOS4-8 & 30234 & 9202 & 661 & 44  \\
noSIDM-8 & 27497 & 5088 & 481 & 41 \\
sDAO-8 & 27037 & 5056 & 475 & 41\\
CDM-6.5 & 77429 & 15690 & 1687 & 178\\
ETHOS4-6.5 & 40175 & 14314 & 1674 & 181 \\
noSIDM-6.5 & 35437 & 8765 & 1283 & 178 \\
sDAO-6.5 & 34390 & 8580 & 1292 & 181 \\
\hline
\end{tabular}
\end{table}

The primary physical intuition from Fig.~\ref{fig:raw_clustering} is a well-known feature of clustering bias, where higher mass objects are clustered more strongly. A second takeaway is the tendency towards a mean, as the galaxy separation increases the strength of clustering decreases, and thus approach the limit of being randomly distributed.. We limit our clustering calculations to a maximum separation of $3.5$ Mpc, which is approximately one tenth of our simulation box size. 

Interestingly, we see that sDAO models have consistently higher clustering values for the minimum mass limit of $10^6 \,{\rm M}_\odot$. This is a result of the ratio of satellite galaxies to central galaxies being greater for sDAO models in this mass limit. We find that for a galaxy selected at a fixed stellar mass, the proportion of the sample that is composed of satellites is around 15 per cent higher in the sDAO model than in CDM. This is due to the fact that central galaxies dominate the CDM mass function which, while also true, is less pronounced in sDAO due to the suppression in the mass function. As a result, a stellar mass selected sample in sDAO contains a higher proportion of satellites (relatively speaking). Figure~\ref{fig:central_ratio} demonstrates the difference in central galaxy abundance discussed above. The fact that the higher clustering is found predominantly for galaxy separations of less than 1 Mpc reinforces our finding that the high clustering values are caused by satellite galaxies. When the minimum mass threshold is raised, there is smaller chance of picking up satellite galaxies to meet the required number density; as a result, the sub-Mpc differences in clustering are no longer observed.

It is difficult to discern more general model differences from this diagram. To condense the clustering information we parametrize individual curves by fitting them to the following functional form:

\begin{figure}
    \centering
    {\includegraphics[width = \columnwidth, height=0.28\textheight]{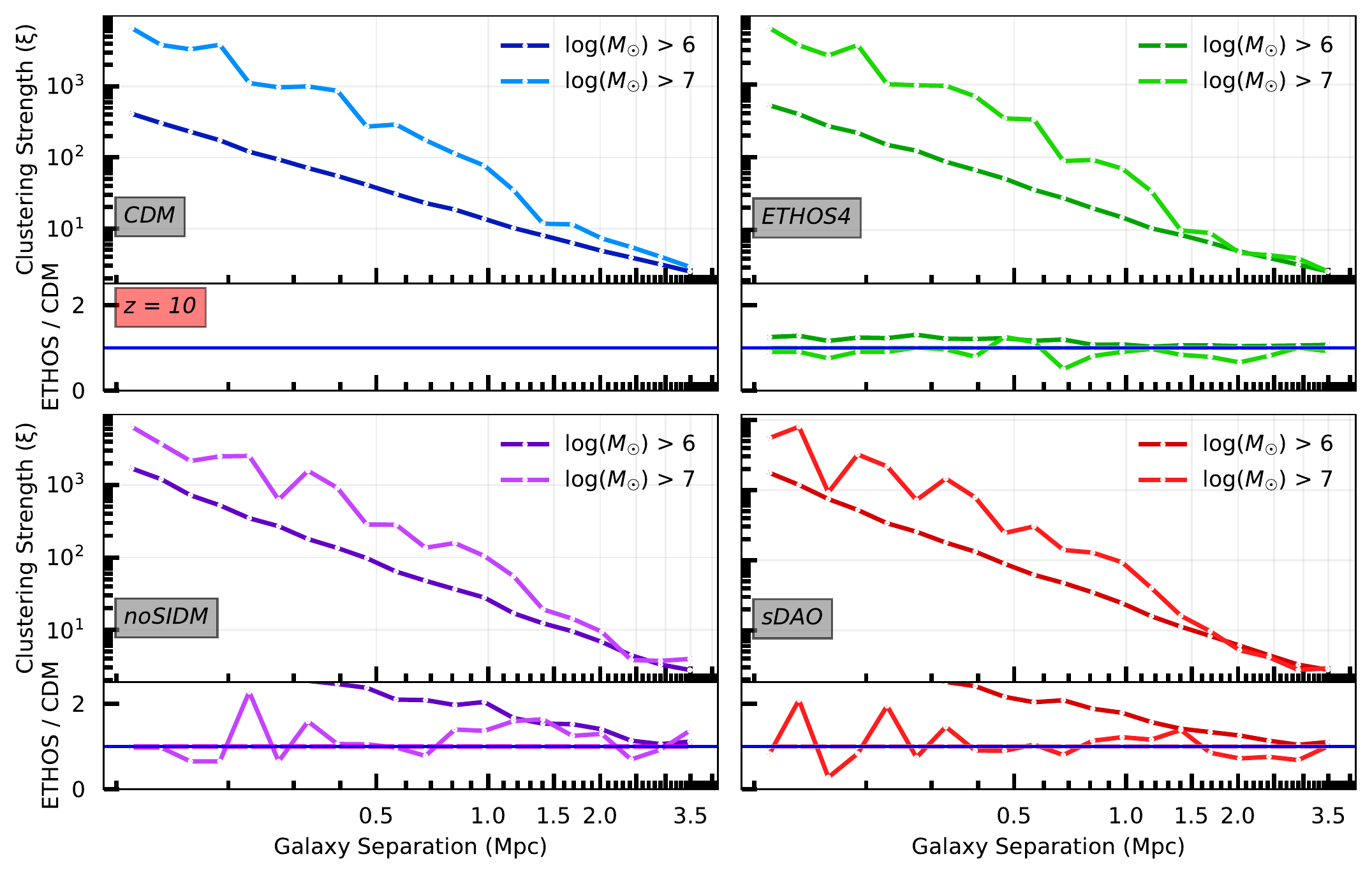}}
    \\
    {\includegraphics[width = \columnwidth, height=0.28\textheight]{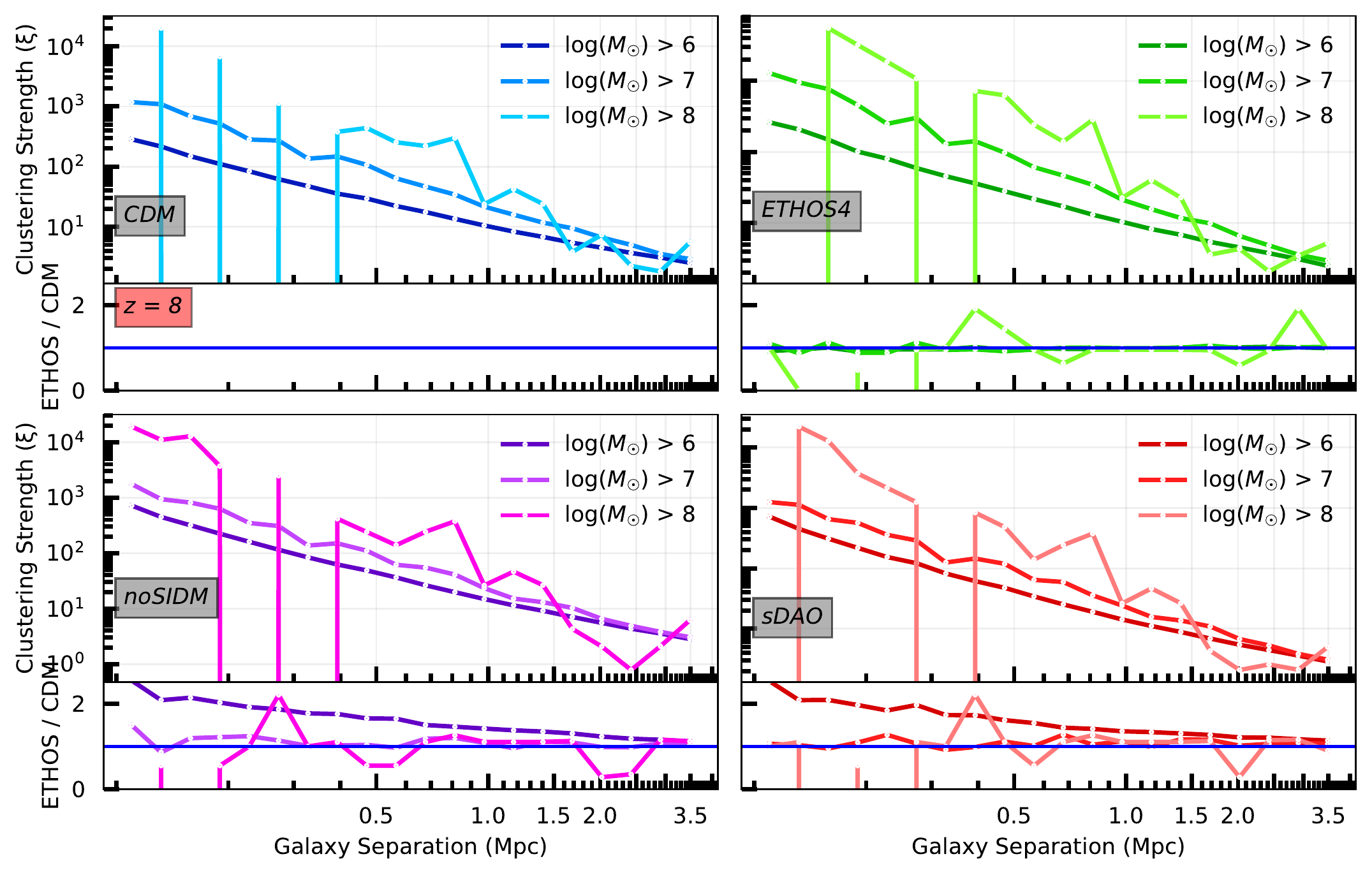}}
    \\
    {\includegraphics[width = \columnwidth, height=0.28\textheight]{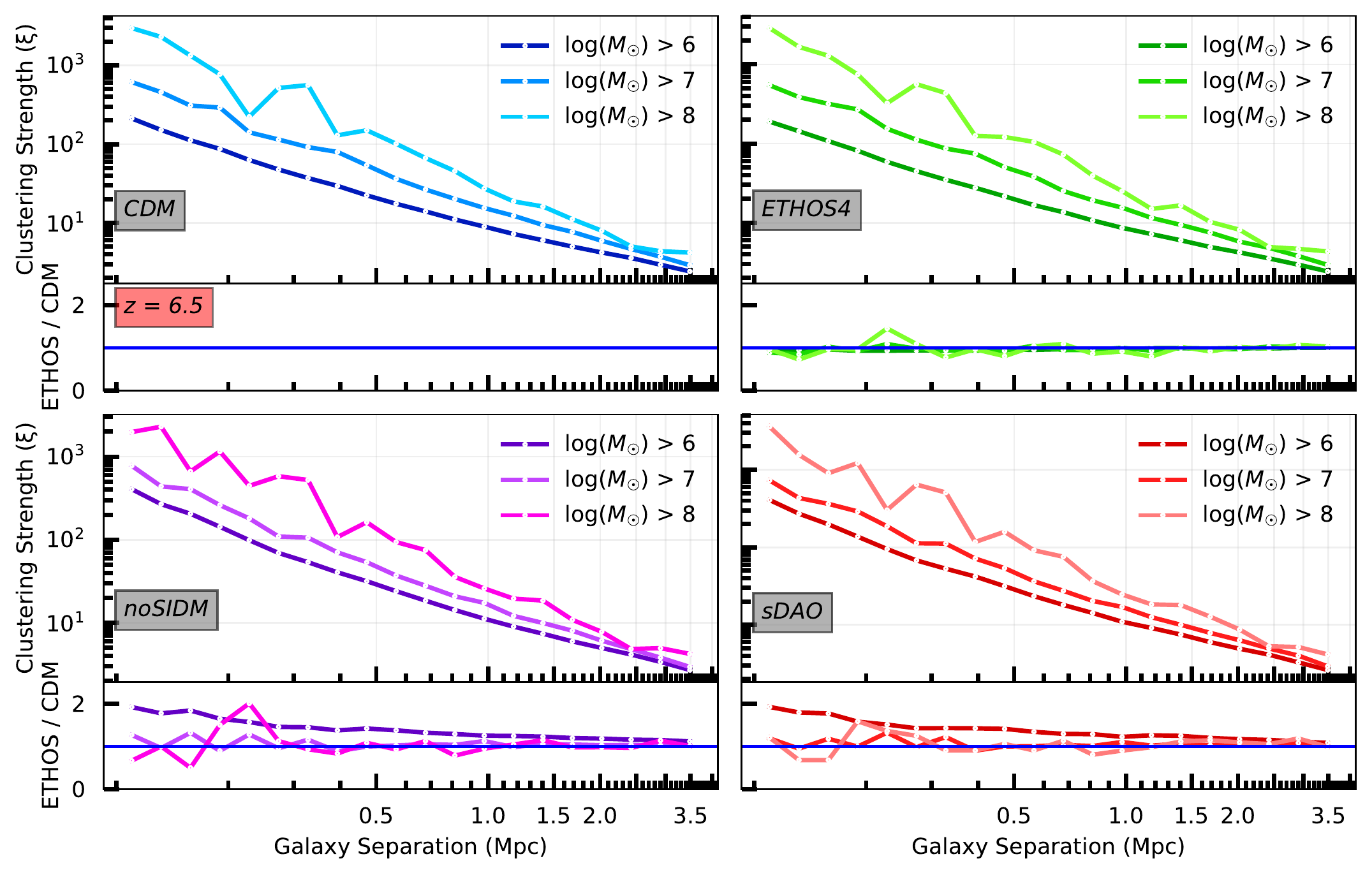}}
    \caption{Galaxy clustering (two-point correlation function) for different models in redshifts 10, 8, and 6.5. The top plots of the panels are the strength of clustering with respect to a random distribution as a function of galaxy separation. The bottom panels show the ratio between the model and CDM. The different models are indicated by different sequential colour maps and plotted in separated labeled panels. Shades of blue, green, magenta, and red, represent CDM, ETHOS4, noSIDM, and sDAO respectively. The clustering of higher mass objects are plotted with lighter colours. }%There is not a clearly distinguishable difference between clustering statistics other than $10^6 {\rm M}_\odot$ minimum mass limit. Difference in that range is explained in the text and Fig.~\ref{fig:central_ratio}}
    \label{fig:raw_clustering}
\end{figure}

\begin{figure}
    \centering
    \includegraphics[width = \columnwidth]{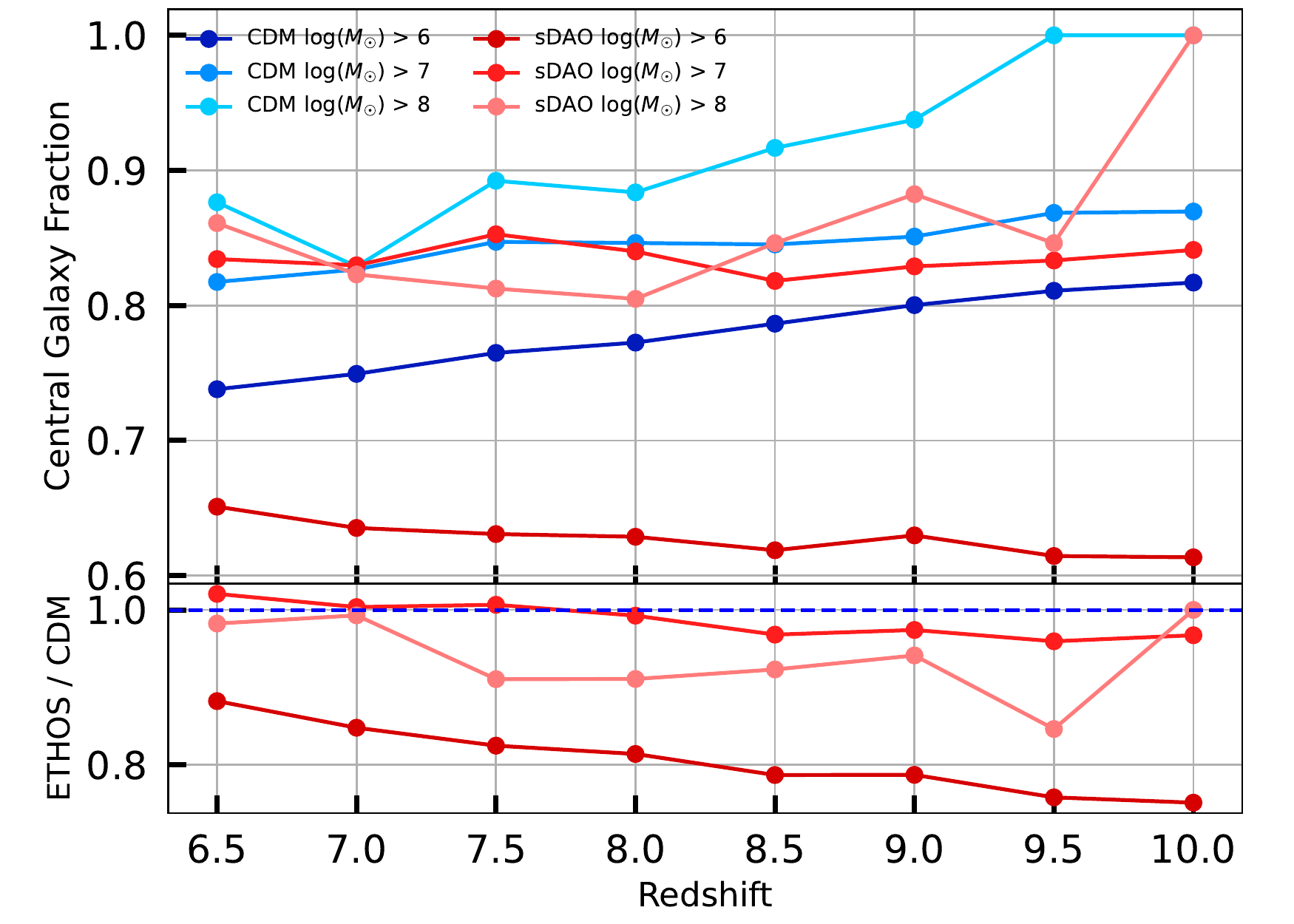}
    \caption{The ratio of number of central galaxies to the total number of galaxies at a particular redshift for CDM (blue) and sDAO (red). The different colour scales show different minimum mass limits, the lighter the colour higher the mass. There is a relatively significant difference on the plots of minimum mass of $10^6\, {\rm M}_\odot$ which explains the consistently high clustering values on Fig.~\ref{fig:raw_clustering}. The ratios on the second plot show that at minimum mass of $10^6 {\rm M}_\odot$, the difference of ratios go up to 20 per cent.}
    \label{fig:central_ratio}
\end{figure}

\begin{equation}
\label{eq:clustering}
    \xi = \left(\frac{r}{r_0}\right)^{-\gamma}.
\end{equation}
We use the {\tt scipy.optimize.curvefit} \citep{SciPy_2020} package to fit Equation~\ref{eq:clustering} to the individual clustering measurements, allowing us to represent the information in Fig.~\ref{fig:raw_clustering} in two dimensions with the variables $r_0$ and $\gamma$. Increasing $r_0$ implies increasing clustering strength, this is because it shows data clustering is equivalent to a random distribution at larger galaxy separations. Increasing $\gamma$ suggests increasing radial dependence of the stregth of clustering. 
\begin{figure*}
    \centering
    {\includegraphics[width = 0.8\textwidth]{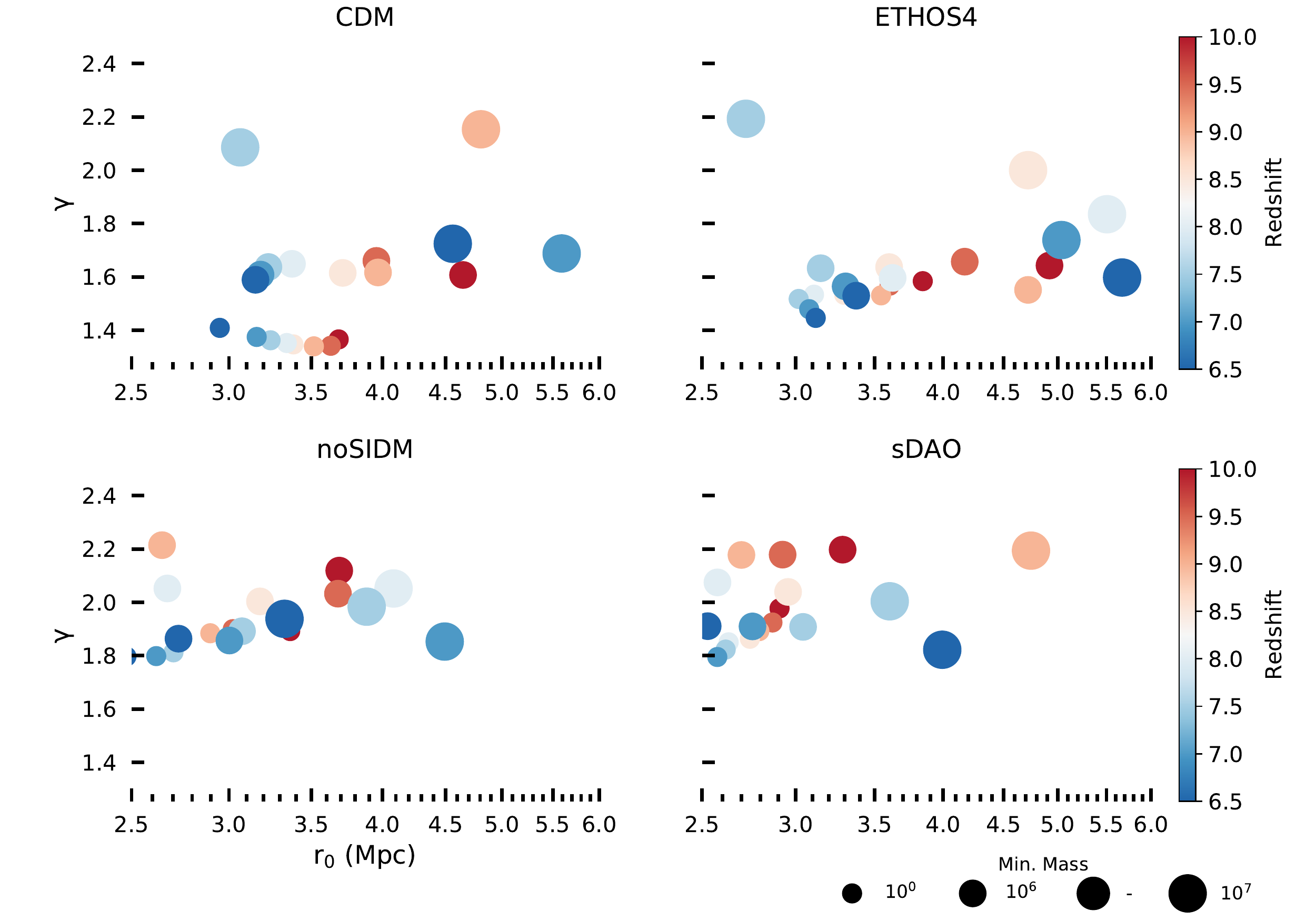}}
    \caption{Condensed representation of the clustering properties of the galaxy population based on a fit to the correlation function. We used the {\tt scipy.optimize.curvefit} to fit the clustering data in Fig.~\ref{fig:raw_clustering} into parameters $r_0$ and $\gamma$ in Eq~\ref{eq:clustering}. Each plot has four dimensions ($x$, $y$, colour, size). Subplots depict different models. The $x$ axes are the $r_0$ (the radii where clustering becomes equal to a random distribution) values of the clustering plots in Mpc. The $y$ axes are the $\gamma$ values of the curve fit and show the radial dependence of clustering plots. The colours of the points represent redshift. They diverge from red to blue going from higher to lower redshift and are shown in colorbar. The different sizes represent different lower mass limits. The legend portrays the specific sizes. The low mass limits are  $10^0$, $10^6$, $10^7$ which means that corresponding plots only include galaxies above those masses. Compared to Fig.~\ref{fig:raw_clustering}, this figure lets us visually distinguish the general differences between the models.}
    \label{fig:corr_processed}
\end{figure*}

The condensed representation of all the clustering data in Fig.~\ref{fig:corr_processed} shows the plane of $\gamma$ and $r_0$. Individual symbols in each panel show the result at different redshifts, while the size of each symbol represents the threshold mass of the galaxy sample used for the clustering measurement. We ignore the case when the stellar mass threshold is $10^8\,{\rm M}_\odot$ as there are too few galaxies to measure the clustering reliably.
 This figure provides an interesting way to perceive the effects of DM-DR and DM-DM interactions. Compared to the less extreme models (CDM and ETHOS4), sDAO-noSIDM have systematically higher values of $\gamma$. This suggests that as the deviation from CDM of the model increases, the radial dependence of galaxy clustering increases; this is driven predominantly by the enhanced one-halo clustering in the sDAO model. Although our statistics are compromised by the limited simulation volume, it suggests that the radial dependence of galaxy clustering could be used as a means to probe models with strong DM-DR and DM-DM interactions in observational data. We note that some of the $r_0$ values are in Fig.~\ref{fig:corr_processed} $> 3.5$ Mpc, which is larger than the maximum separation we measure the clustering to. This is because the fit in Equation~\ref{eq:clustering} tries to extrapolate the value.

We have mentioned previously how the number of galaxies included in the clustering calculation affects the result. Table \ref{tab:number_of_galaxies} shows that the less extreme models CDM and ETHOS4 have consistently more galaxies for a particular mass threshold and redshift combinations. Although the corresponding galaxy counts for all the models agree within an order of magnitude, there is more than 20 per cent difference between the two model classes. Such a difference could result in less accurate curve fitting and higher values of $\gamma$ for extreme models.

To eliminate the possibility of a numerical cause for the radial dependence of clustering, we have re-performed the clustering calculations by choosing the same amount of galaxies from each model i.e. computing the clustering at {\it fixed number density}. For a particular combination of redshift and low mass limit, we determined which one out of the 4 models had the least number of galaxies, then we chose the least amount in descending stellar mass order from other models. For example, at the low mass threshold of $10^7 \,{\rm M}_\odot$ and redshift 6.5, the noSIDM model has 1283 galaxies while CDM, ETHOS4, sDAO respectively have 1687, 1674, 1292. For that redshift, we only select the 1283 most luminous galaxies from each model., consequently,

After doing the same curve fitting and plotting the normalized clustering calculations, the less extreme models had relatively higher radial dependence ($\gamma$) values than in Fig.~\ref{fig:corr_processed}. Specifically, CDM, at the low mass threshold of $10^0 \,{\rm M}_\odot$, had a mean $\gamma$ value of $\sim 1.71$ with a standard deviation of $\sim 0.0475$ compared to sDAO with a mean and standard of deviation of $\sim 1.86$ and $\sim 0.0648$. CDM still has a difference of three standard deviations. In addition, at the low mass threshold of $10^6 \,{\rm M}_\odot$ which will be observable with lensed JWST survey, CDM had a mean and standard deviation of $\sim 1.86$ and $\sim 0.0891$ while sDAO had a mean $\sim 2.05$ with standard deviation $\sim 0.126$. From a statistical perspective, clustering behaviour of different models should be distinguishable even in the normalized case because either distributions are more than two standard deviations away (CDM) or have a kurtosis lower than a normal distribution. This reinforces our notion that the stronger radial clustering relative to CDM is a physical outcome of the ETHOS models.

However, choosing the same amount of galaxies made the $\gamma$ values of CDM and ETHOS4 overlap with more extreme models in contrast to a clear difference in Fig.~\ref{fig:corr_processed}. That said, we are still limited by simulation volume and, consequently, the decreased accuracy of curve fitting at high redshift. Thus, despite clustering analysis being a possible way to constrain alternative dark matter models, we need larger simulations to further quantify the feasibility of this method.

\section{Conclusions}
\label{sec:Conclusions}

Observations of the early universe will be at the forefront of space-based programmes such as the {\it James Webb Space Telescope} (JWST). The high redshift universe is a particularly fertile environment for testing the predictions of different models of DM. To this end, we compared cosmological simulations of galaxy formation of the well known CDM model with models in which a massive DM particle couples with a massless relativistic particle through a massive mediator. The alternative models were a part of the ETHOS ({\it Effective Theory of Structure Formation}) framework and are characterized by their initial power spectrum given by the strength of the DM-dark-radiation interaction in the early Universe, and also by the strength of DM-DM interactions. Specifically, the ETHOS4 model is a fine-tuned model with parameters that would optimally alleviate CDM problems. The sDAO and noSIDM are models with very strong DM relativistic interactions that would result in prominent galactic scale differences. The sDAO and noSIDM models are used as a test case of the effect of DM-self interactions compared to relativistic interactions: the sDAO model has self-interactions enabled while noSIDM does not. The simulations we performed have $2 \times 1024^3$ particles and were run over a box of $(36.2\, {\rm Mpc})^3$. Our goal was to investigate the observable consequences of each DM model on the galactic population at the redshift range relevant to missions like the JWST.

We began by analyzing the DM halo populations of the different simulations. Although haloes are not directly observable, they are the main building blocks which provide general intuition about the simulated universe. The abundance of DM haloes as a function of mass shows the primary phenomenological differences between each of the models. All models predict largely the same number of high mass objects (halo mass $>10^{10} {\rm M}_\odot$). The alternative models suppress the number of dwarf scale structures and demonstrate their purpose of alleviating CDM problems. While ETHOS4 is more similar to CDM down to lower halo mass compared to sDAO models, the nature of the cutoff of structure to even smaller scales is much sharper (Fig.~\ref{fig:halo_mass_functions}). These effects can therefore be traced all the way back to the initial conditions.

We next contrasted the differences of galaxies in an analogous way to halos (Fig.~\ref{fig:stellar_mass_functions}). The distributions of galaxies with respect to their mass show that there is a mapping from the halo distribution. While the number of brightest galaxies, above $10^8\, {\rm M}_\odot$, are indistinguishable between the models, in the mass range of $10^7-10^8\, {\rm M}_\odot$, the sDAO models have half as many galaxies as ETHOS4 or CDM. Still, such a difference is insufficient to distinguish between models using JWST due to theoretical uncertainties of $\pm 0.5$ dex. However, with strong gravitational lensing, JWST may be able to observe galaxies between $10^6-10^7\, {\rm M}_\odot$ in which difference of stellar mass functions of CDM and extreme models are around half an order of magnitude. This suggests JWST augmented by a strong gravitational lensing survey may be a promising way to constrain DM models using the faint end of the stellar mass functions. We still noted that sDAO models had DM-DR interactions that were 1 million times stronger compared to ETHOS4 which does not translate to the relatively small amplitude of the differences on galactic scales (only a difference around half order of magnitude).

To further understand the connection between galaxies and haloes in these models we analyzed the stellar-to-halo mass ratios and galaxy occupancy fraction of  haloes with respect to the mass ranges (Figs.~\ref{fig:halo_luminosity} and \ref{fig:halo_star_ratio}). Both the median values and the general distribution of galaxy-to-halo mass ratio of all the models are nearly identical. The efficiency of galaxy formation decreases with halo mass, and reaches a minimum at $10^9\, {\rm M}_\odot$ in all models. 

In addition to the stellar-to-halo mass ratio, we have also quantified the fraction of haloes that host a galaxy to the total number of haloes in a particular mass range -- the so-called `luminous fraction' of DM haloes as a function of mass. We performed this calculation with two qualifications for defining a galaxy: one with haloes hosting any luminous component, and one with haloes with at least 10 star particles, corresponding to $\sim 10^6\, {\rm M}_\odot$. The behaviour of the models are very similar in both cases. Such observed similarity of models in two statistical analysis of galaxy and halo relations enables us to conclude that the galaxy-halo relation within the context of our simulations is more strongly related to the galaxy formation physics (whose parameters were fixed across all models) rather than the particle physics of the models.

We next compared the evolution of star formation rates in each of these models (Fig.~\ref{fig:sfr_densityl}). The ETHOS galaxies that ended up with same stellar mass as their CDM counterparts at redshift 6.5 have rapid star formation from the redshifts 10 to 6.5. While the total star formation rate of the universe is higher in CDM at redshift 10, the other models catch up by redshift 6.5. The distribution of contribution of mass ranges to the total as well as the absolute total value of star formation rate become more similar between the models as time passes, see Fig.~\ref{fig:total_sfr_breakdown}. Nevertheless, while the 60 percent of star formation of CDM comes from dwarf scale galaxies at redshifts 10 and 9.5, such contribution is less than 50 per cent for ETHOS models. %Overall, although we had found some differences in the simulations, the differences DAOs were theoretically anticipated to cause were not in the scale large enough to be used as observational constraints.

Finally, we considered the clustering of galaxies as a means of probing exotic DM models (Fig.~\ref{fig:raw_clustering}), a statistic that may be feasible to compute in the era of JWST. We  evaluated the clustering strength of galaxies as a function of galaxy separation for different models, different stellar mass thresholds, and at different redshifts. We find that as the amount of the relativistic coupling increases, the radial dependence of clustering increases and the overall strength of clustering decreases (Fig.~\ref{fig:corr_processed}). We then verify that this increased radial dependence is indeed a physical effect which is related to the relative fraction of satellite and central galaxies in a given sample (Fig.~\ref{fig:central_ratio}). Of course it is challenging to measure clustering at  $z \geq 6$, but JWST will provide by far the best opportunity to measure the galaxy correlation function in the high redshift universe. Future ETHOS simulations performed in larger volumes will enable a more robust comparison of the clustering strength of these alternative DM models relative to CDM.

A persistent feature in all the analyses we have performed is that the sDAO and noSIDM models are nearly identical, both qualitatively and quantitatively. This demonstrates that at high redshifts, due to the low probability of collisions between DM particles, the DM-DM interactions are not an effective factor in influencing the properties of galaxies.

A caveat to any work of this kind is that a coherent understanding of galaxy formation physics is still lacking. As we are still uncertain about the actual baryonic processes operating in galaxies, our results have to be considered primarily within the framework of models like IllustrisTNG. Nevertheless, the kinds of differences we have measured in Section~\ref{sec:clustering} are fundamentally outcomes of differences in the halo mass function, and are thus more likely to be representative of the underlying particle physics differences. We have also specifically chosen models to maximize galactic scale differences. Less extreme models will undoubtedly deviate to a smaller extent; nevertheless, the present work provides a sense for the maximum deviation from CDM expected in the class of exotic DM models considered here.

\section*{Acknowledgements}

We thank the referee for making a number of helpful suggestions that have improved the clarity of our work. AK acknowledges support from the Harvard PRISE Fellowship. SB is supported by the UK Research and Innovation (UKRI) Future Leaders Fellowship [grant number MR/V023381/1]. MRL acknowledges support by a Grant of Excellence from the Icelandic Research Fund (grant number 206930). CP acknowledges support by the European Research Council under ERC-AdG grant PICOGAL-101019746. JZ acknowledges support by a Grant of Excellence from the Icelandic Research fund (grant number 206930).

%%%%%%%%%%%%%%%%%%%%%%%%%%%%%%%%%%%%%%%%%%%%%%%%%%
\section*{Data Availability}

The data presented in the figures are available upon request from the corresponding author.
 


\bibliographystyle{mnras}
\bibliography{references} % if your bibtex file is called example.bib

%\begin{thebibliography}{99}
%\end{thebibliography}

%%%%%%%%%%%%%%%%%%%%%%%%%%%%%%%%%%%%%%%%%%%%%%%%%%

%%%%%%%%%%%%%%%%% APPENDICES %%%%%%%%%%%%%%%%%%%%%

%\appendix

% \section{Some extra material}

% If you want to present additional material which would interrupt the flow of the main paper,
% it can be placed in an Appendix which appears after the list of references.

%%%%%%%%%%%%%%%%%%%%%%%%%%%%%%%%%%%%%%%%%%%%%%%%%%

% Don't change these lines
\bsp	% typesetting comment
\label{lastpage}
\end{document}